\documentclass[aps,pra,groupedaddress,twocolumn,longbibliography]{revtex4}

\usepackage[pdftex]{graphicx}
\usepackage{epstopdf}
\usepackage{amsmath,amssymb,amsfonts,hyperref,physics}
\usepackage{natbib}
\usepackage{float}
\usepackage{xcolor}
\usepackage{soul}
\usepackage{chngcntr} 
\hypersetup{
	colorlinks,
	linkcolor={red!50!black},
	citecolor={blue!50!black},
	urlcolor={blue!80!black}
}

\definecolor{jasper}{rgb}{0.88, 0.23, 0.24}

\begin{document}

\title{Human perceptual decision making of nonequilibrium fluctuations}

\author{Ayb\"uke Durmaz$^{1}\footnote{Contributing author: \url{adurmaz@sissa.it}}$, Yonathan Sarmiento$^{1,2}\footnote{Contributing author: \url{ysarmien@sissa.it}}$ , Gianfranco Fortunato$^{1}$, Debraj Das$^{2}$,\\ Mathew E. Diamond$^{1}$, Domenica Bueti$^{1}$\footnote{Corresponding author: \url{bueti@sissa.it}}, and \'Edgar Rold\'an$^{2}$\footnote{Corresponding author: \url{edgar@ictp.it}}\medskip}

\affiliation{$^1$International School for Advanced Studies (SISSA), Via Bonomea 265, 34136 Trieste, Italy\\
$^2$ICTP $-$ The Abdus Salam International Center for Theoretical Physics, Strada Costiera 11, 34151 Trieste, Italy}


\begin{abstract}
    Perceptual decision-making frequently requires accumulating noisy sensory inputs as evidence in order to execute rapid, reliable choices. Neural recordings from the lateral intra-parietal area in humans and primates performing perceptual decision-making tasks highlighted evidence accumulation mechanisms, often modeled as drifted diffusions. In those experiments, participants were exposed to computer-generated randomly fluctuating stimuli whose motion is not linked to any physical phenomenon. To better characterize the statistical processes underlying human decision-making, we performed experiments where human participants visualized fluctuations of physical nonequilibrium stationary states, and we analyzed responses in the context of stochastic thermodynamics. A total of forty fiveparticipants viewed hundreds of movies of a particle endowed with drifted Brownian dynamics and were tasked with judging the motion as leftward or rightward in a quick and reliable manner. Overall, the results uncover fundamental performance limits, consistent with recently established thermodynamic trade-offs (uncertainty relations, TURs) involving speed, accuracy, and dissipation; specifically, lower rates of entropy production lead to longer decision times.
    Moreover, to achieve a given level of observed accuracy, participants require more time than predicted by Wald’s optimal sequential probability ratio test, indicating suboptimal integration of available information. In view of such suboptimality, we develop an alternative account equipped with non-Markovian evidence integration with a memory time constant, and find tight fits. Our results suggest that humans adapt their memory relaxation time to the rate of dissipation of the observed phenomenon, favouring memory over momentary evidence for effective decisions in scenarios where stimuli are far from equilibrium. Furthermore, we  identify the effects of the environmental stability on decision-making performance and memory  by comparing the results of the two sets of experiments: blocked (stationary) versus intermixed (non-stationary) conditions. Our study illustrates that perceptual psychophysics using stimuli rooted in nonequilibrium physical processes provides a robust platform for understanding how the human brain makes decisions on stochastic information inputs.
\end{abstract} 

\maketitle

\section{Introduction}
\label{perceptual_decision_making}

A key function of the brain is to process noisy sensory data to make fast and accurate decisions ~\cite{gold2007neural}. In the simplest scenario, a binary decision between two competing hypotheses $H_1$ and $H_2$, Wald's sequential probability ratio test (SPRT)~\cite{wald1945sequential}, rooted in probability theory~\cite{tartakovsky2014sequential}, is a plausible universal model for optimal decision-making in diverse fields of natural sciences, as it outlines how sensory information should be integrated over time to maximize the accuracy of the decision (statistics~\cite{wald1948optimum}, biophysics~\cite{kobayashi2010implementation,siggia2013decisions,desponds2020mechanism}, neuroscience~\cite{ratcliff2016diffusion, ratcliff2008diffusion,forstmann2016sequential}). The SPRT accumulates ongoing evidence by computing the log-likelihood probability ratio between $H_1$ and $H_2$ until the weight of evidence towards any of the hypotheses exceeds a predefined threshold value. In neuroscience, SPRT has been applied to drift-diffusion models (DDM)~\cite{ratcliff2008diffusion} where a decision is taken as soon as the accumulated log-likelihood ratio, which fluctuates instantaneously but has a net drift, reaches one of the two thresholds. Combining the SPRT and DDM paradigms gives a physical first-passage time problem in a bounded interval, where the boundary that is first reached ($H_1$ or $H_2$), and the exit time from the interval correspond to the decision and the decision time, respectively~\cite{ratcliff2015modeling}.

Various tasks with rhesus monkeys~\cite{salzman1990cortical,shadlen2001neural,roitman2002response}, rats~\cite{rigosa2017dye}, and humans~\cite{ratcliff2008diffusion} yield actual decision-making outcomes close to the optimal performance predicted by SPRT, giving credence to sequential hypothesis testing models~\cite{gold2007neural,gerstner2014neuronal}.  
One widely used task involves judging the net motion within a random dot motion (RDM) display on a screen. In studies that used this task, rhesus monkeys are trained to respond with a saccade to the screen side that matched the perceived net motion (e.g. saccade to the right (left) for the net motion rightward (leftward))~\cite{shadlen2001neural}. Across trials, the level of coherence among moving dots in the selected direction is manipulated, making a given trial easy or hard. The activity of spatially selective neuronal populations in the lateral intra-parietal area (LIP) is found to be correlated with the decision of monkeys. 
More specifically,  the activity in LIP neurons coding for the chosen spatial location gradually increases and peaks around the onset of the saccade. Furthermore, the slope of increase is correlated with the coherence of the stimuli, that is, the higher the coherence, the steeper the increase and suppression~\cite{roitman2002response, shadlen2001neural, gold2007neural, gerstner2014neuronal, salzman1990cortical}. 
These findings reveal neuronal operations that correspond to accumulation of evidence toward a threshold, as proposed by the DDM (SPRT) framework.  

The nature of evidence accumulation cannot be considered separately from the properties of the evidence that is acquired.  Recent research highlights the relationship between complex spatiotemporal fluctuations in stimuli and decision outcomes~\cite{gerstner2014neuronal,park2016spatiotemporal,fard2022stochastic, insabato2014influence, wimmer2015sensory, urai2017pupil, zylberberg2012construction, kiani2008bounded, peixoto2021decoding}. For instance, one study demonstrates that participants consistently account for detailed spatiotemporal noise patterns within stimuli, rather than relying solely on the average noise within a trial~\cite{fard2022stochastic}. Similarly, a series of experimental works find that the predictive power of decision-making models is improved when incorporating data from stochastic  input trajectories rather than only average-based, nonspecific noise~\cite{insabato2014influence, brunton2013rats}. Thus, while it is established that instantaneous, not only average, properties of dynamical systems are weighted in decision making, physical systems in nonequilibrium produce a distinct form of sensory evidence characterized by fluctuations that exhibit time irreversibility (e.g. net currents of matter and energy) and energy dissipation ~\cite{fang2019nonequilibrium}. It is not yet clear which computational framework best characterizes how human observers acquire evidence from such systems.

To answer this question, we carry out visual perception experiments where participants judge movies of a driven Brownian particle described by a Langevin equation with a few nonequilibrium degrees of freedom. Data analysis, grounded in the framework of stochastic thermodynamics, establishes a robust link between participants' behavioral responses and the input parameters of the physical Brownian process.  The study consists of Experiments A and B. Experiment A enables us to test a theoretical model of binary decision-making against data of human participants, quantifying the extent to which their decisions deviate from optimality defined by the model; Experiment B enables us to address the effect of non-stationary properties of the environment on optimality; the control experiments enable to address the variance in the data due to non-decisional factors. Our approach combines multiple disciplines -- behavioral psychophysics, computational decision-making frameworks, and stochastic thermodynamics to provide a means to refine decision-making models beyond the SPRT and to uncover fundamental thermodynamic constraints~\cite{roldan2015decision} that shape judgments of nonequilibrium stationary stimuli. Furthermore, the findings offer insights into the relationship between the processing of nonequilibrium fluctuations and optimality (or suboptimality) in perceptual decision making. 
The paper is structured as follows: Sec.~\ref{sprt_on_times_arrow} briefly reviews Wald's SPRT with regard to its application to assessing the arrow of time of a 1D drift-diffusion process. Sec.~\ref{nonequi_percep_experiment} gives details of both Experiments A and B. Sec.~\ref{data_analysis} provides a statistical analysis of the findings from Experiment A and compares them with the predictions of Wald's SPRT. Sec.~\ref{sec:exII}  presents the findings of Experiment B. Sec.~\ref{sec:TUR}  employs the empirical data to test the validity of dissipation-time thermodynamic trade-offs that have been identified in stochastic thermodynamics for Markovian systems. Finally, Sec.~\ref{sec:EIM}  introduces a non-Markovian Evidence Integration Model (EIM) to incorporate the effects of memory in perceptual decision-making and gives a statistical comparison between SPRT and EIM. Sec.~\ref{discussion} consists of a discussion of all findings. 
Detailed descriptions of the experimental setup, data acquisition and analysis, mathematical derivations, and statistical tests including cross-validation are provided in the appendices.

\section{Sequential probability ratio test on the direction of time's arrow}
\label{sprt_on_times_arrow}

Consider a time series of snapshots characterized by a nonequilibrium stationary stochastic process $X_t$. By observing this evidence, one must make a decision $D_{\mathrm{dec}}$ by choosing between two competing hypotheses, 
each corresponding to a physical model of the process, e.g., $H_1 = 1$ and $H_2 =-1$. 
When judging such a stochastic process repeatedly across a sequence of independent trials, how should one process the evidence to {\em optimally} achieve the minimal decision time (averaged across trials) for a given prescribed level of accuracy?
The first rigorous answer to this question emerges from Wald's SPRT: the optimal decision time $T_{\rm dec}$ to select $D_{\mathrm{dec}} = 1$ $(D_{\mathrm{dec}} = -1)$ is when the cumulative log-likelihood ratio ${\mathcal{L}}_t= \ln [ {\mathcal{P}} \qty( X_{[0,t] }  | H_1) / {\mathcal{P}} \qty( X_{[0,t] }  | H_2) ] $ 
first exceeds (first falls below) a fixed decision threshold $L_{+}$ $(L_{-})$ determined by the prescribed accuracy~\cite{wald1945sequential,wald1948optimum}.
Here, the decision accuracies are the conditional probabilities $\alpha_{+} = P ( D_{{\mathrm{dec}}} = 1 | H_1 )$ and $\alpha_{-} = P ( D_{{\mathrm{dec}}} = -1 | H_2 )$, and denote, respectively, the following.  
The probability that the decision $D_{{\mathrm{dec}}} = 1$ ($D_{{\mathrm{dec}}} = -1$) when the hypothesis $H_1$ ($H_2$) is indeed true. On the other hand, the quantity ${\mathcal{P}} \qty( X_{[0,t] }  | H_1)$ $({\mathcal{P}} \qty( X_{[0,t] }  | H_2) )$ denotes the conditional probability of observing the trajectory $X_{[0,t]}$ given hypothesis $H_1$ ($H_2$) is true.
For a broad class of continuous stochastic processes, 
the decision thresholds ensuring symmetric accuracies $\alpha_{\pm} = \alpha$ are given by $L_{\pm} = \pm \lambda$ with $\lambda = \ln[\alpha/(1-\alpha)]$~\cite{tartakovsky2014sequential}. 
Wald's SPRT is thus a first-passage-time problem for ${\mathcal{L}}_t$ with the decision time 
$T_{\mathrm{dec}} = \inf \qty{ t \geq 0 | {\mathcal{L}}_t \notin \qty(-\lambda, \lambda) }$ given by the first-exit time from  the interval $(-\lambda,\lambda)$.  

In thermodynamics, an enduring question is how the time irreversibility of nonequilibrium processes gives rise to entropy production in the form of, e.g., heat dissipation. Consider the paradigmatic 1D drift-diffusion model~(1DDM) for an overdamped Brownian particle in one dimension 
$    \dot{X}_t = v + \sqrt{2 D} \xi_t,$ 
 where $X_t$ denotes the particle's position at time $t$; $v$ and~$D$ are the drift velocity  and the diffusion constant, respectively. 
The quantity $\xi_t$ is a zero-mean Gaussian white noise with autocorrelation $\expval{\xi_t \xi_s} = \delta(t-s)$. 
Suppose the task of an SPRT is to decide as soon as possible, achieving some prescribed average accuracy across trials, whether an observed stochastic path
$X_{[0,t]}$  (with initial position $X_0 = 0$ and periodic boundary conditions) is produced by a 1DDM with positive speed $v>0$ (hypothesis $H_1$) or instead by a time-reversed
1DDM with negative speed $v<0$ (hypothesis $H_2$). 
The log-likelihood ratio associated with such SPRT reads 
\begin{equation}
    {\mathcal{L}}_t = \ln \left[ \frac{{\mathcal{P}} \qty( X_{[0,t] }  | H_1)  }{ {\mathcal{P}} \qty( X_{[0,t] }  | H_2)}  \right] =\left(\frac{ v}{D}\right) X_t, \label{eq:log1}
\end{equation} and coincides with the stochastic entropy production~\cite{seifert2005entropy} associated with the stochastic trajectory~$X_{[0,t]}$~\cite{roldan2015decision}, which for the 1DDM is proportional to the position (see Appendix~\ref{app:theo}).   Note that we have ${\mathcal{P}}(X_{[0,t]}|H_2) = {\mathcal{P}}(\widetilde{X}_{[0,t]}|H_1)$, where ${\mathcal{P}}(\widetilde{X}_{[0,t]}|H_1)$ is the probability of observing the time-reversed trajectory $\widetilde{X}_{[0,t]}$ under  hypothesis $H_1$ ($v>0$) true. Therefore $\mathcal{L}_t$ for this problem encodes the degree of time-reversal asymmetry associated with the statistics of the model $H_1$.  
These relations imply that the SPRT  evidence accumulation rate is given by
\begin{equation}
    \langle\dot{\mathcal{L}}_t \rangle=  \frac{v \langle\dot{X}_t\rangle}{D} = \frac{v^2}{D} .
\end{equation}
Interestingly, we have   $ \langle\dot{\mathcal{L}}_t \rangle=\Sigma$, where $\Sigma$ is the steady-state rate of entropy production associated with the 1DDM (in units of the Boltzmann constant $k_{\rm B}$). The latter is given by the steady-state average power divided by $k_{\rm B}\mathsf{T}$ with $\mathsf{T}$ being the temperature of the environment, i.e.,
\begin{equation}
    \Sigma = \frac{\langle\dot{W}\rangle}{k_{\rm B} \mathsf{T} }= \frac{v^2}{D}.
    \label{eq:3}
\end{equation}
In Eq.~\eqref{eq:3}, we use $\langle\dot{W}\rangle=\gamma v^2$ as the power dissipated by a constant friction force $\gamma v$ and Einstein's relation $D=k_{\rm B}\mathsf{T}/\gamma$. 

Furthermore, the SPRT on the direction of time`s arrow maps into the problem of first exit time for $X_t$ to reach any of the two decision thresholds located in $L$ and $-L$ with $L = \lambda D/v = D \ln[\alpha_{\mathrm{SPRT}}/(1-\alpha_{\mathrm{SPRT}})]/v$. 
Consequently, the decision time is given by $   T_{\mathrm{dec}} = \inf \qty{ t \geq 0 | X_t \notin \qty(-L, L) } $, whose mean value retrieved from the first-passage theory~\cite{redner2001guide}  reads $ \expval{T_{\mathrm{dec}}}_{\mathrm{SPRT}}  = (L/v) \tanh\left( vL/2D\right)$.  Using Eq.~\eqref{eq:3}  
 and introducing the dimensionless P\'{e}clet number
\begin{equation}
    \mathrm{Pe} \equiv \frac{ v L}{D},
\end{equation}
we obtain the SPRT mean decision time  
\begin{align}
\expval{T_{\mathrm{dec}}}_{\mathrm{SPRT}} = \frac{\mathrm{Pe}}{\Sigma} \tanh \qty(\frac{\mathrm{Pe}}{2}).
\label{eq:meanT_wald}
\end{align}
Moreover,  we obtain that the predicted accuracy set by the absorption probability $\alpha_{\mathrm{SPRT}} =P(X_{T_{\rm dec}}=L)$ is equal to
\begin{align}
\alpha_{\mathrm{SPRT}} = \frac{1}{1+ \exp(-\mathrm{Pe})} .
\label{eq:SPRT_accuracy}
\end{align}
Note that Eqs.~\eqref{eq:meanT_wald} and~\eqref{eq:SPRT_accuracy} reveal that the mean decision time and the accuracy in the SPRT depend only on two parameters, $\Sigma$ and $\text{Pe}$. 
See Appendix~\ref{app:theo} for more details of the derivation of Eqs.~(\ref{eq:log1}-\ref{eq:SPRT_accuracy}). 

Let us now consider a suboptimal decision maker that collects stochastic evidence from the DDM and takes decisions based on a first-exit problem of  $\mathcal{J}_t$,  a functional of $X_{[0,t]}$ that may be in general different from the log-likelihood ratio (e.g. $\mathcal{J}_t=\int_0^t {\rm{d}}s f(s)X_s$, with $f$ a prescribed real function). Let $(-L_J,L_J)$ be the interval chosen such that the accuracy of the decision maker equals that of the SPRT [Eq.~\eqref{eq:SPRT_accuracy}], leading to the mean decision time $\langle T_{\rm dec}\rangle$. Under such assumptions, the SPRT's optimality implies the relation $\langle T_{\rm dec}\rangle\geq \expval{T_{\mathrm{dec}}}_{\mathrm{SPRT}} $, which can be generalized to a broad class of stochastic processes~\cite{tartakovsky2014sequential}. For the 1DDM, this relation together with Eq.~\eqref{eq:meanT_wald}  implies that
\begin{equation}
    \Sigma \expval{T_{\mathrm{dec}}} \geq 
    \mathrm{Pe} \, \tanh \qty(\frac{\mathrm{Pe}}{2}) ,
    \label{eq:TURdt}
\end{equation}
i.e. an inverse relation between dissipation (entropy production) and (mean decision) time; see Appendix~\ref{app:theo} for further details. Eq.~\eqref{eq:TURdt} belongs to the emerging collection of quantitative predictions of thermodynamic constraints on the accuracy of nonequilibrium currents in the burgeoning field of thermodynamic uncertainty relations (TURs)~\cite{Barato2015,Gingrich2016,Falasco2020,Horowitz2020}. A fundamental lower limit to the mean decision time of the form of Eq.~\eqref{eq:TURdt} was first conjectured for Markovian processes~\cite{roldan2015decision} and later proved and generalized to generic nonequilibrium stationary processes~\cite{Neri2021}, but here tested in experimental scenarios for the first time.

\begin{figure*}
\includegraphics[scale=0.76]{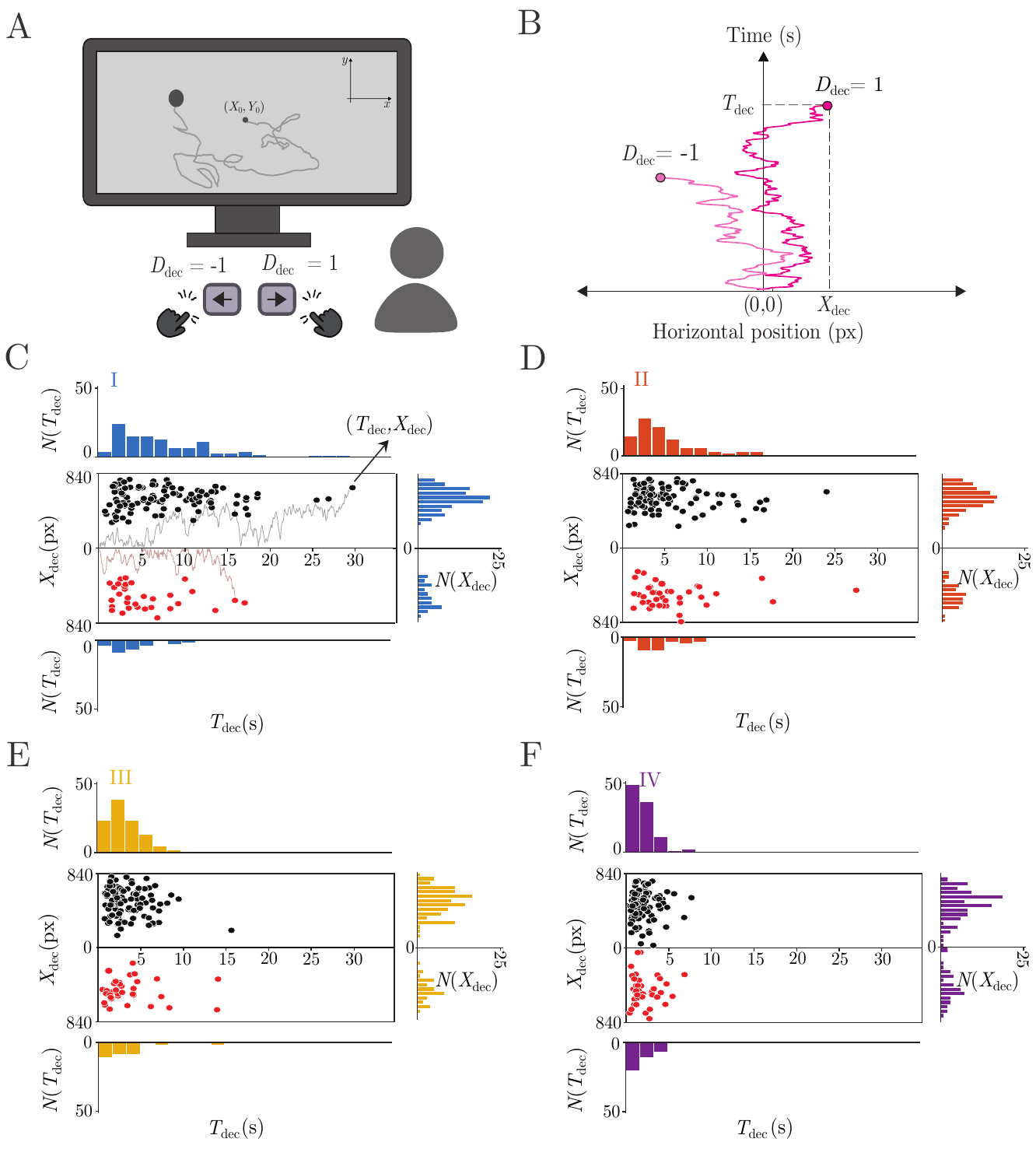}
\caption{{\bf Sketch of the experimental setup and data collection.} (A) Participants are instructed to judge the net motion direction (left versus right) of a moving disk displayed on a computer screen by pressing the right (left) arrow key on a keyboard. 
The snapshots of the disk's motion are generated from stochastic simulations described in the main text.  $D$.  
(B) Example traces of $x$-position over time are shown in fuchsia. For each trial, we collect the decision time ($T_{\rm dec}$, in seconds), the position of the disk along the $x$-axis ($X_{\rm dec}$) and $y$-axis ($Y_{\rm dec}$, not illustrated here), in pixels, and the decision outcome  ($D_{\rm dec} = 1$ for right, or $D_{\rm dec} = -1$ for left). We also store the trajectories of the disk up to the decision time (gray line in panel A). 
(C-F) Experimental data collected from a typical participant across the four experimental conditions in Experiment A, each characterized by a selected value of $v$ and~$D$. For each condition, the plots in the middle show $X_{\rm dec}$ as a function of $T_{\rm dec}$. The black and red dots indicate correct and incorrect decisions, respectively. Histograms are obtained from empirical counts of  $T_{\rm dec}$  for correct (top) and incorrect decisions (bottom) and for  $X_{\rm dec}$  for correct (top) and incorrect (bottom) decisions. In C-F, we pool the statistics of trials with positive and negative drift for both correct and incorrect decisions.
}
\label{Fig1}
\end{figure*}

\section{Nonequilibrium perceptual experiments at work}
\label{nonequi_percep_experiment}

In two experiments, we ask 45 healthy human participants to decide the net direction of a stochastic visual stimulus moving with a given drift velocity $v$ and a given diffusion coefficient $D$. More precisely, the two sets of experiments are designed as follows: Experiment~A  (\textit{N}=21 participants) provides a first exploration of the degree of optimality in decisions by looking at four conditions (labeled as I-IV) with a fixed ratio of $v/D$ but with different values of the entropy production rate $\Sigma=v^2/D$ . In A, a single session contains 50 sequential trials with a fixed entropy production rate. Experiment~B  (\textit{N}=24 participants) investigates optimality in a changing environment by (a) a  intermixing entropy production rate across trials within a session, and (b) a broader parameter space (combinations of $v$ and $D$), consisting of nine different conditions, with a wider range of the ratio $v/D$ and $\Sigma$ (labeled as a-i). For more details, see Appendix \ref{sec:addexp}. 

Fig.~\ref{Fig1}A sketches the experimental setup with a computer screen, on which movies of a dark gray disk are displayed on a light gray background. In each trial, the trajectory of the disk's center of mass ($X_t$, $Y_t$) is obtained by simulating the nonequilibrium dynamics of an overdamped Brownian particle in two dimensions (see Appendix~\ref{materials}). The corresponding dynamics is set by the Langevin equation $\dot{X}_{t} = Hv + \sqrt{2D}\xi_{t}$, $\dot{Y}_{t} = \sqrt{2D}\eta_{t}$. Here, $v>0$ is the drift velocity along the $x$-axis, and $D>0$ is the diffusion coefficient, which we take equal along the $x$ and $y$ axes. The dichotomous random variable $H = \{1,-1\}$ sets the left/right direction of the drift and in each trial is set randomly to -1 or 1 with equal probability. The quantities $\xi_{t}$ and $\eta_{t}$ are independent Gaussian white noises, each with zero mean and satisfying correlations $\langle\xi_{t}\xi_{s}\rangle =\langle \eta_{t}\eta_{s}\rangle =\delta(t-s)$, $\langle \xi_t \eta_s\rangle=0$ for all $t,s$. 
Such simulations mimic the fluctuating motion of an overdamped Brownian particle dragged at a constant net speed $v$ in a Newtonian fluid with viscosity $\gamma =k_{\rm B}\mathsf{T} / D$, where $k_{\rm B}$ is Boltzmann's constant and $\mathsf{T}$ is the temperature of the fluid. Thus, our experimental setup serves as a minimal yet flexible platform to study human perception of nonequilibria.

Participants are asked to decide whether the direction of the disk is rightward ($D_{\rm dec} = 1$) or leftward ($D_{\rm dec} = -1$) (Fig.~\ref{Fig1}A). The free decision time is defined as $T_{\rm dec}$.
After participants complete 20 training trials to familiarize themselves with the task, experimental data are collected. In both training and experimental sessions, participants receive feedback on the accuracy of each response. 
No knowledge of the underlying dynamics of the disk is provided to the participants.
For each trial and each participant, relevant quantities are recorded: (i) the true hypothesis (ground truth), i.e., the actual direction of the net drift $H=\{1,-1\}$; (ii) the decision outcome $D_{\rm dec} = \{1,-1\}$; (iii) the decision time $T_{\rm dec}$; (iv) the position of the disk along the $x$-axis of the screen at the decision time, $X_{\rm dec}=X(t = T_{\rm{dec}})$ which corresponds to \textit{decision thresholds} (position of the disk along the $y$-axis at decision time was recorded but proved not to be critical to any analysis); and (v) the full stochastic trajectory $X_{[0,T_{\rm{dec}}]}$ of the disk up to~$T_{\rm{dec}}$  (see Fig.~\ref{Fig1}B). 

To explore the decision consistency across participants given a particular stimulus, we present to each participant the same set of movies of stochastic trajectories, but randomize the presentation order of distinct movies across participants (see Appendix~\ref{sec:var_in_decision_threshold}).

\section{Experiment A}
\label{data_analysis}

Here, we explore four conditions with varying entropy production rates (denoted as I, II, III, and IV) in a block design, where the trials of the same condition are grouped together. The entropy production rate $\Sigma = v^2/D$ increases monotonically from condition I to condition IV, with a fixed ratio $v/D$ (see Table~\ref{tab:1} in Appendix~\ref{materials} for the values of $v$ and $D$ in each condition). 

Figs.~\ref{Fig1}C-F display the data collected from a participant in the four experimental conditions, each comprising $150$ independent trials. For each condition, the plots in the middle show $X_{\rm dec}$ as a function of $T_{\rm dec}$. The black and red dots indicate correct and incorrect decisions, respectively. The histograms of $T_{\rm dec}$ (top and bottom) and $X_{\rm dec}$ (on the right) are plotted separately for correct and incorrect decisions.

\begin{figure*}%
\centering
 \includegraphics[scale=0.38]{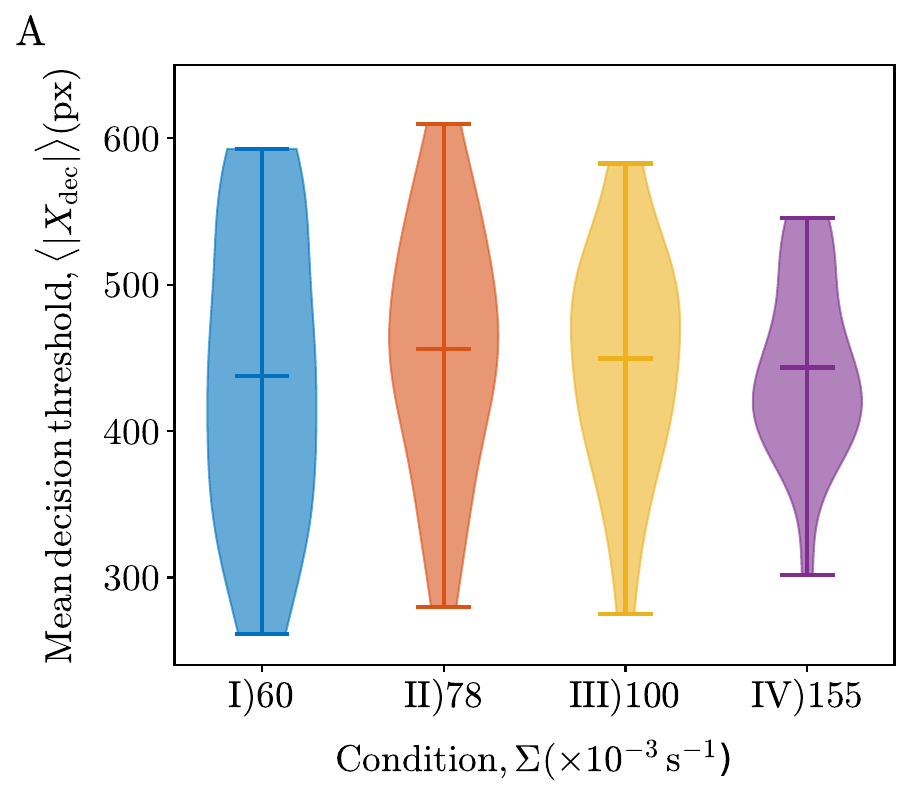}
  \includegraphics[scale=0.38]{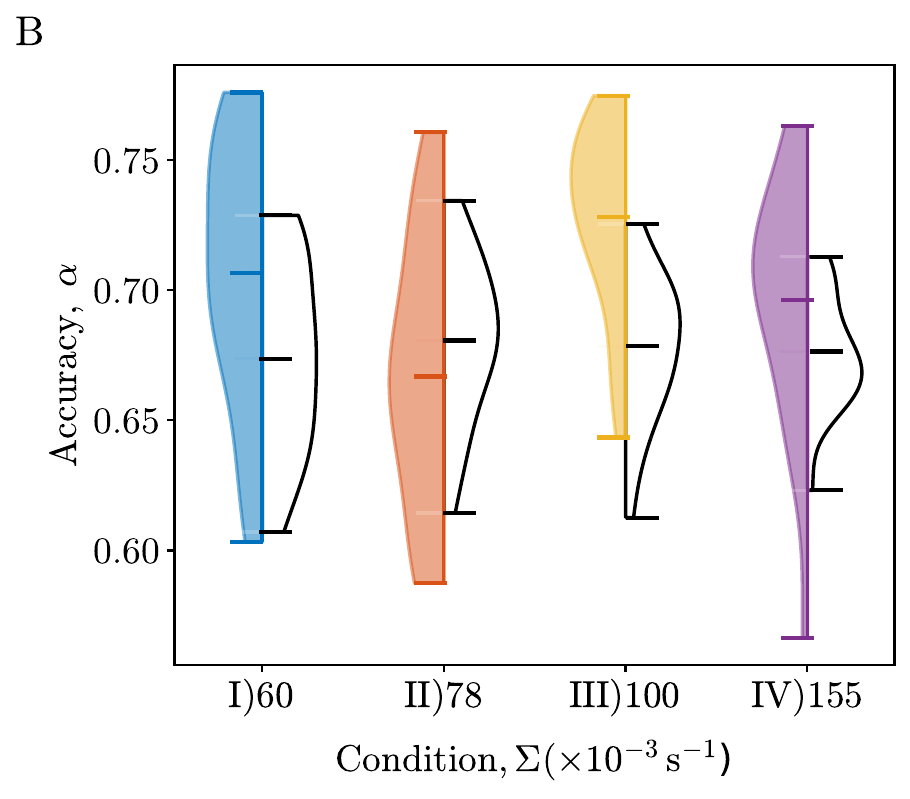}
   \includegraphics[scale=0.38]{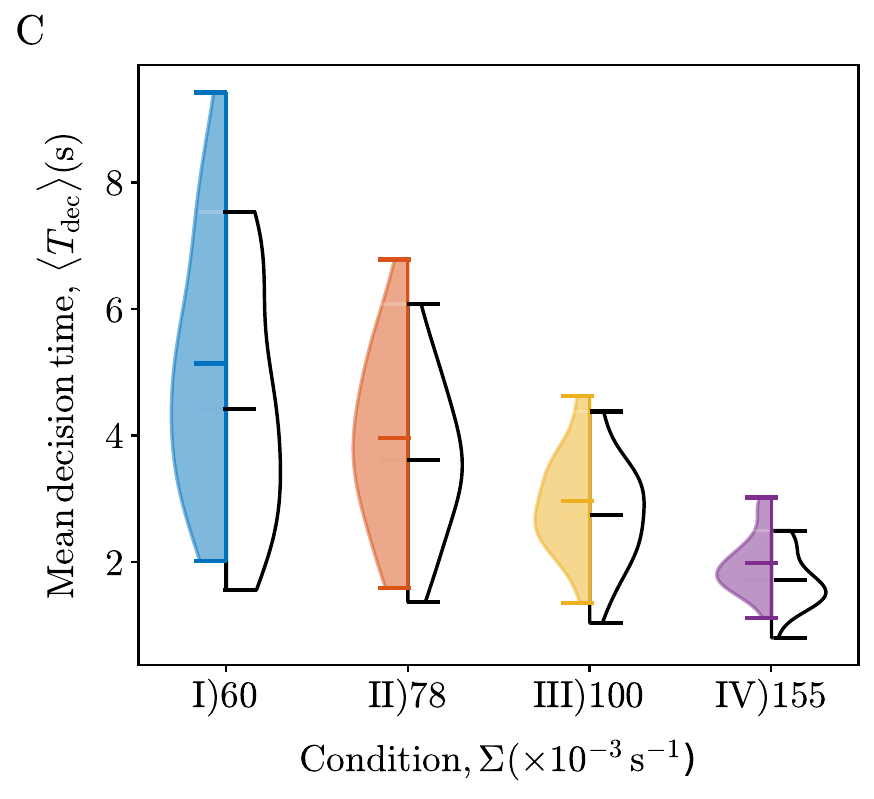}
\caption{{\bf Experiment A, statistics of the participants' mean values of decision-making quantities as a function of experimental conditions labeled by entropy production rate $\Sigma$.} (A) Distribution of mean decision threshold $\langle \vert X_{\rm dec} \vert \rangle$,  (B) accuracy $P(D_{\rm dec}=H)$, and (C) mean decision time $\langle T_{\rm dec}\rangle$. Experimental (filled) violin plots are obtained by collecting the statistics of each of the $N=21$ participants' mean decision threshold amplitude (A), accuracy (B), and mean decision time (C).  Plots are obtained by averaging over all trials for every participant. 
The empty violin plots in (B, C) illustrate the distribution of the theoretical predictions from SPRT with thresholds set at  $\pm\langle \vert X_{\rm dec} \vert \rangle$ obtained from experiment for each participant. Central horizontal lines show the respective means. SPRT predictions of mean decision time and accuracy are obtained by plugging  P\'eclet number and entropy production rate in Eqs.~\eqref{eq:meanT_wald} and~\eqref{eq:SPRT_accuracy}, respectively.}    
\label{fig3}
\end{figure*}

To inspect the effects of the entropy production rate $\Sigma$ on decision-making at the ensemble level, we focus on the mean decision threshold, mean accuracy, and mean decision time, averaged over all trials and over all participants under four experimental conditions (plotted as colored violin plots in Fig.~\ref{fig3}).

First, we test the changes in the mean decision threshold $\langle X_{\rm dec}\rangle $ across conditions with a linear mixed effect (LME) model ($X_{\rm dec}$ model formula: $\langle \textit{X}_{\mathrm{dec}} \rangle \sim \text{condition} + (1 \mid \text{id})$\footnote{The notation represents a linear mixed-effects model. Here, $\langle \textit{X}_{\mathrm{dec}} \rangle$ denotes the dependent variable (e.g., decision variable), modeled as a function of a predictor (condition) and a random intercept $(1 \mid \text{id})$, which accounts for individual variability by allowing each participant (id) to have their own baseline shift. If multiple predictors are included, interactions can also be examined, as in $\langle \textit{X}_{\mathrm{dec}} \rangle \sim \text{\textit{v}} * \text{\textit{D}} + (1 \mid \text{id})$, where $\text{\textit{v}} * \text{\textit{D}}$ explicitly models both main effects and their interaction}, marginal $R^2$:  0.007, conditional $R^2$: 0.83).  Type III ANOVA on model estimates reveal no main effect of experimental conditions (see Appendix~\ref{sec:lin_mix_eff}: Decision thresholds for details on LME and ANOVA). This result suggests that participants do not change their mean decision thresholds with varying stimuli dynamics across experimental conditions (Fig.~\ref{fig3}A).

Second, we test whether the participants' mean accuracy varied across experimental conditions with respect to the typical value of $\sim 70\%$ correct  (Fig.~\ref{fig3}B), again using an LME (accuracy model formula: $\alpha$ $ \sim \text{condition} + (1 \mid \text{id})$, marginal $R^2$:  0.25, conditional $R^2$: 0.74). Type III ANOVA on model estimates reveals a statistically significant effect of the experimental conditions on accuracy with $F_{3,60}$ = 25.94, $p <$ 0.001,  $\eta^2$ = 0.56 (see Appendix~\ref{sec:lin_mix_eff}: Accuracies ($\alpha$) for post-hoc multiple comparisons).

Then, we analyze how the mean decision time $\langle T_{\rm dec}\rangle $ changes across conditions with LME ($T_{\rm dec}$ model formula: $\langle T_{\rm dec} \rangle$ $ \sim \text{condition} + (1 \mid \text{id})$, marginal $R^2$:  0.52, conditional $R^2$: 0.92). Type III ANOVA on model estimates reveals a main effect of experimental conditions on $T_{\rm dec}$ with $F_{3,60}$ = 170.36, $ p <$ 0.001, $\eta^2$ = 0.89.  Post-hoc analysis reveals that as $\Sigma$ increases from condition I to IV, $T_{\rm dec}$ decreases (all  $t_{60}$ $>$ 3.89,  $p <$ 0.0015). See Appendix~\ref{sec:lin_mix_eff}: Decision times for details.

These results are in line with the prediction of previous theoretical work~\cite{roldan2015decision} that it will take less time on average to decide on the direction of time's arrow when observing a process that is far from equilibrium; here, this corresponds to highly irreversible disk motion (Fig.~\ref{fig3}C).

Further, in Figs.~\ref{fig3}B and~\ref{fig3}C we compare the theoretical predictions for the SPRT mean accuracy [Eq.~\eqref{eq:SPRT_accuracy}] and mean decision time [Eq.~\eqref{eq:meanT_wald}] (plotted as empty violin plots) with the experimental values, (plotted as filled violin plots). To calculate the SPRT predictions of accuracy as in Eq.~\eqref{eq:SPRT_accuracy}, we consider symmetric accuracies and symmetric thresholds located at $\pm L$, with $L=\langle \vert X_{\rm dec}\vert\rangle $ set at the ensemble average of the absolute value of the experimental decision thresholds; this follows from the fact that the P\'eclet number ${\rm Pe} = vL/D$. Then, chi-square (${\chi}^{2}$) goodness of fit tests are performed for each condition to determine to what extent the distribution of the accuracies predicted from SPRT is close to the distribution of the experimental accuracy data.  ${\chi_c}$ values for conditions I, II, III, and IV are 0.68, 0.42, 1.62, and 0.43, respectively. These values indicate the standard deviation of the difference in theoretical and experimental data of accuracy in comparison to a distribution with zero mean.
Similarly,  chi-square (${\chi}^{2}$) goodness of fit performed to compare the predicted and experimental decision time distributions gives $\chi_c$ values of 1.16, 0.62, 0.83, and 1.19 for Conditions I, II, III and IV, respectively. These results indicate that the SPRT can also describe the mean decision times in all experimental conditions (see the Appendix~\ref{chi_sup} for additional details).
\section{Experiment B}
\label{sec:exII}

Experiment A demonstrates that participants' performance across conditions aligns with the predictions of SPRT. Specifically, both the mean accuracy and the mean decision times are well approximated by the model. However, some degree of suboptimality is evident, with real decision times consistently exceeding the time predictions of SPRT. 
In Experiment B, we hypothesize that suboptimality is based partially on the participants' uncertainty about the statistics of the environment. We predict that, under conditions of greater uncertainty, responses will become more distant from optimality. To test this, we extend the range of the stimulus parameters ($v$ and $D$), covering a larger parameter space. Second, we shift from the block design used in Experiment A to an intermixed design, where trials of different conditions are randomly presented. 
Following the hypothesis above, we speculate that in the block design participants adjust to stimulus patterns more effectively, leading to faster and more accurate decisions. In contrast, intermixed designs will require within-trial adjustment, causing slower response times and reduced accuracy ~\cite{los1999identifying}. 

In Experiment B the physical equations governing the stimulus motion, the experimental task, and the instructions provided to participants are identical to A. A new, nonoverlapping pool of participants is engaged for Experiment B. Motion is defined by nine conditions, each corresponding to a unique combination of $v$ and $D$ values, forming a 3 $\times$ 3 grid (see Appendixes~\ref{materials} and ~\ref{sec:addexp} for details on the experimental design). In Experiment A, both $v$ and $D$ increase with $\Sigma$ from condition I to condition IV. To disentangle the distinct effects of $v$ and $D$ on decision-making, Experiment B manipulates these parameters independently. In addition, the trial movies for the nine conditions are presented in a randomized, intermixed design.

\begin{figure*}%
\centering
 \includegraphics[scale=0.45]{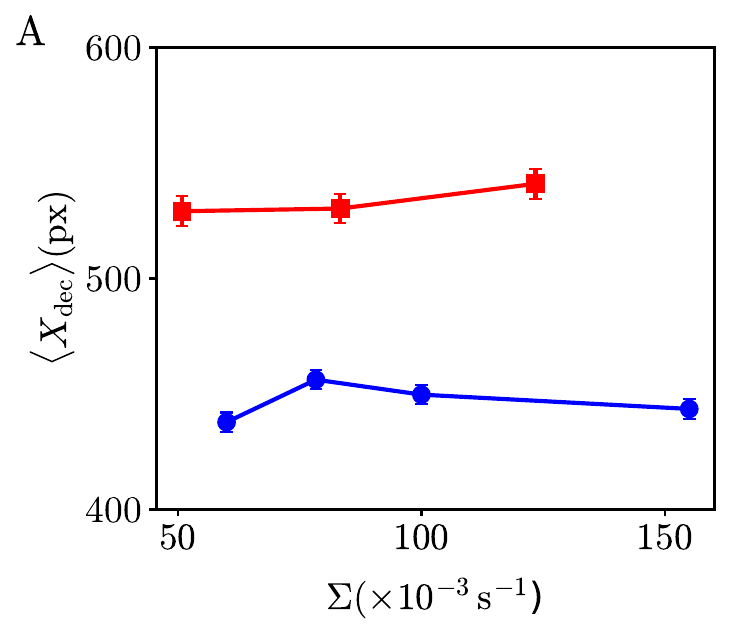}
  \includegraphics[scale=0.45]{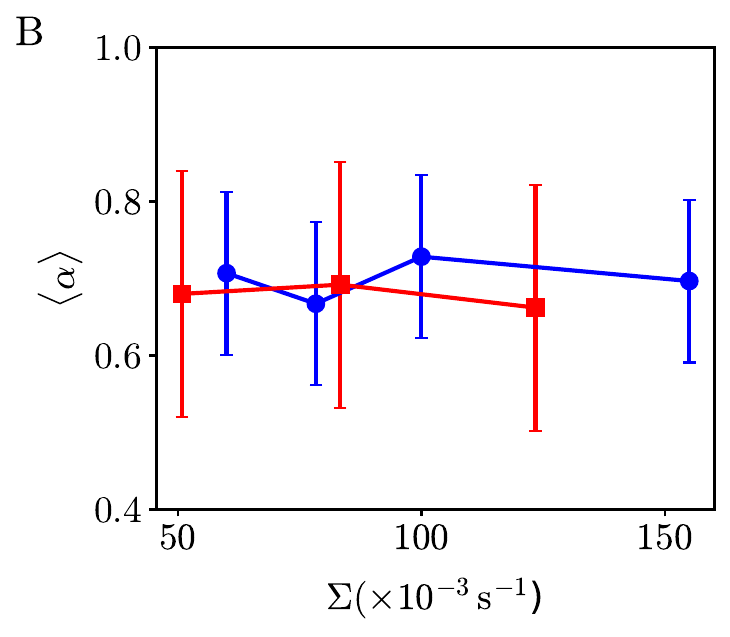}
   \includegraphics[scale=0.45]{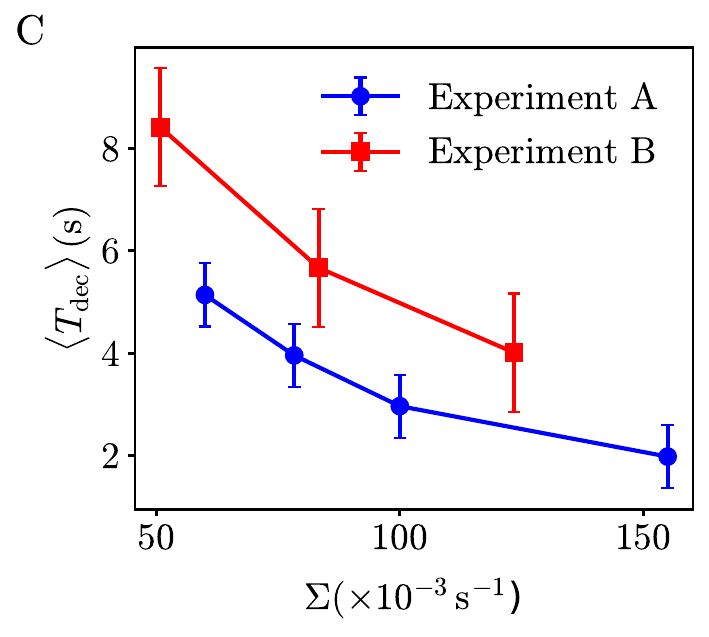}

\caption{{\bf Comparison of the decision-making quantities between two experiments: Experiment A (blue circles) vs Experiment B (red squares).} The panels show decision-making parameters averaged over all  participants as a function of entropy production rate  $\Sigma$.  Note that the conditions from the two experiments are comparable since they have similar $\Sigma$ and fixed Pe. (A) Mean decision thresholds $\langle |X_{\rm dec}|\rangle$ in Experiment B are larger than in Experiment A for all values of $\Sigma$, indicating participants collected more evidence on average in this experiment. (B) Mean accuracy as a function of $\Sigma$. Statistical tests reveal no significant difference between mean accuracies of the two experiments. (C) Mean decision times as a function of $\Sigma$. An inverse proportionality relation between $\langle T_{\rm dec}\rangle$ and $\Sigma$ as predicted by SPRT is observed in both experiments. However, the decision times are delayed in Experiment B. Error bars are standard error of the mean (SEM). The parameters $v$ and $D$ of the data shown in the panels can be found in Table~\ref{tab:1} for Experiment A and in Fig.~\ref{FigPrmSpcTable} (diagonal elements) for Experiment B.  } 
\label{figComp1and2}
\end{figure*}

\subsection{Effect of tuning to the stimuli statistics on optimality}

 \par

The three diagonal conditions (see Fig.~\ref{FigPrmSpcTable}) are comparable with the four conditions of Experiment A, as they are designed to have a fixed $v/D$ ratio with almost identical parameter values. Therefore, in Fig.~\ref{figComp1and2} we compare these three diagonal conditions of Experiment B (red squares) with the four conditions of Experiment A (blue circles). Wilcoxon rank sum tests compare decision metrics (decision thresholds, accuracy, decision times) between the two experiments (Fig.~\ref{figComp1and2}A). Participants in Experiment B have on average larger decision thresholds compared to Experiment A (\textit{W} = 108, \textit{p} = 0.0011, \textit{RBC} = -0.286), suggesting that they collect more data to cope with the unpredictable nature of the environment. Similarly, the mean decision times for Experiment B conditions are delayed compared to Experiment A, as seen in Fig~\ref{figComp1and2}C  (\textit{W} = 35, \textit{p} $<$ 0.0001, \textit{RBC} = -0.431). However, even if they spend more time on average, the accuracy does not significantly increase, \textit{W} = 324, \textit{p} = 0.143, \textit{RBC} = 0.143 (see Fig~\ref{figComp1and2}B). Overall, these results indicate that the less optimal decision outcomes in Experiment B can be explained by a lack of tuning to the stimulus statistics due to the intermixed-design.

\subsection{Exploring the parameter space}

\begin{figure*}%
\centering
 \includegraphics[scale=0.38]{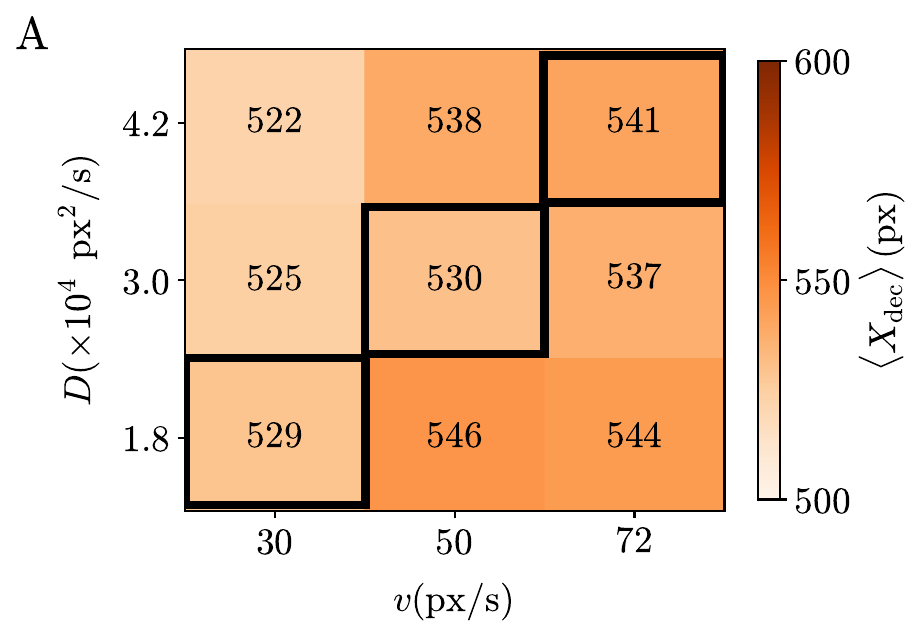}
  \includegraphics[scale=0.38]{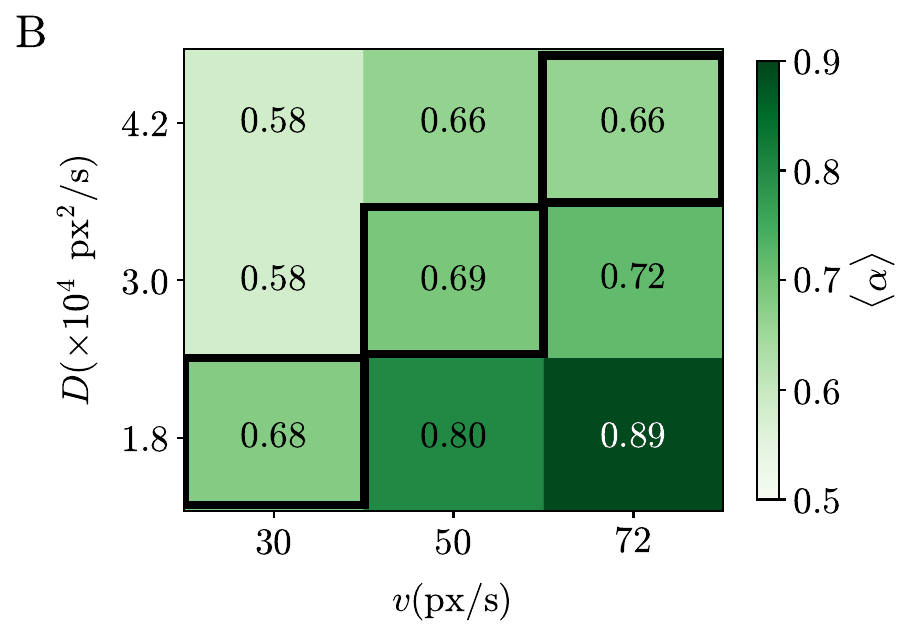}
    \includegraphics[scale=0.38]{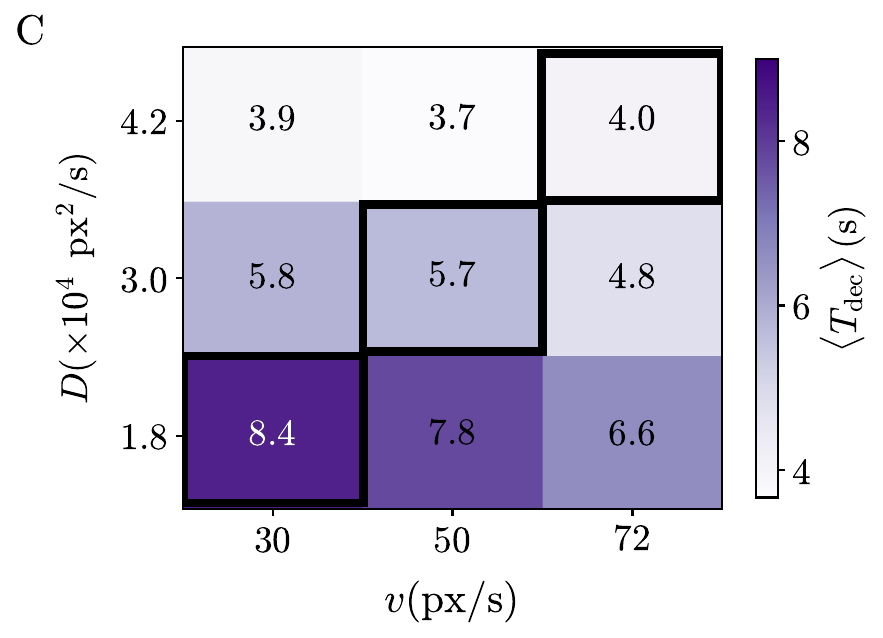}

\caption{{\bf Statistics of the participants’ mean values of decision-making quantities across parameter space in Experiment B.} Each panel represents $3 \times 3$ grid of parameter space in Experiment B, $v$ depicted in $x$-axis and $D$ depicted in $y$-axis with each cell representing a combination of these two parameters. The cells show the numerical value of (A) mean decision thresholds, $\langle |X_{\rm dec}|\rangle$; (B) mean accuracy, $\langle \alpha \rangle$  and (C) mean decision times, $\langle T_{\rm dec}\rangle$. The color bars represent the corresponding numerical value of the decision-making quantity. } 
\label{figPEvsRT}
\end{figure*}

We can exploit the 3 $\times$ 3 grid of parameter space to test the effects of $v$ and $D$ separately on decision-making quantities, which is not possible in Experiment A as these parameters are constrained to have $v/D$ fixed yet different $\Sigma=v^2/D$. First, we test participants' mean decision thresholds (Fig~\ref{figPEvsRT}A) with factors of $v$ and $D$ with LME ($X_{\rm dec}$ model formula: $\langle \textit{X}_{\mathrm{dec}} \rangle \sim \text{\textit{v}} * \text{\textit{D}} + (1 \mid \text{id})$, marginal $R^2$:  0.014, conditional $R^2$: 0.93). 
ANOVA type-III analysis performed on the obtained model reveals a significant main effect of $v$ with $F_{2,184}$ = 14.79, $p <$ 0.001,  $\eta^2$ = 0.14, and $D$ with $F_{2,184}$ = 5.14, $p <$ 0.01,  $\eta^2$ = 0.05; but no interaction  $F_{4,184}$ = 0.77, $p =$ 0.55. 
Post-hoc multiple comparisons reveal that low $v$ conditions are significantly narrower than medium ( $t_{184}$ = 4.38,  $p =$ 0.0001) and high $v$ ( $t_{184}$ = 4.98, $p <$ 0.0001) conditions; and low $D$ conditions are wider than medium $D$ conditions ($t_{184}$ = -3.21, $p =$ 0.0048) (see Appendix~\ref{sec:lin_mix_eff} for details). 

Second, we test accuracy values (Fig~\ref{figPEvsRT}B) with LME ($\alpha$ model formula: $\alpha \sim \text{\textit{v}} * \text{\textit{D}} + (1 \mid \text{id})$, marginal $R^2$: 0.84, conditional $R^2$: 0.93). ANOVA type-III analysis performed on the obtained model accuracy reveals main effects of $v$ with $F_{2,184}$ = 564.85, $ p <$ 0.001, $\eta^2$ = 0.86; and $D$ with $F_{2,184}$ = 707.54, $ p <$ 0.001, $\eta^2$ = 0.88, and their interaction with $F_{4,184}$ = 40.98, $ p <$ 0.001, $\eta^2$ = 0.47. Higher values in $v$ (all $t_{184}$ $>$ 3.190,  $p <$ 0.005 except one pair) and lower values in $D$ (all $t_{184}$ $>$ 3.644,  $p <$ 0.001 except one pair) is associated with an accuracy increase (see Appendix~\ref{sec:lin_mix_eff} for details), as Eq.~\ref{eq:SPRT_accuracy} suggests. 

Similarly,  $T_{\rm dec}$ values are tested (Fig~\ref{figPEvsRT}C) with LME ($T_{\rm dec}$ model formula: $\langle \textit{T}_{\mathrm{dec}} \rangle \sim \text{\textit{v}} * \text{\textit{D}} + (1 \mid \text{id})$, marginal $R^2$:  0.59, conditional $R^2$: 0.96). ANOVA type-III analysis performed on the obtained model $T_{\rm dec}$ reveals main effects of $v$ ($F_{2,184}$ = 22.10, $ p <$ 0.001, $\eta^2$ = 0.19), and $D$ ($F_{2,184}$ = 1551.72, $ p <$ 0.001, $\eta^2$ = 0.94), and their interaction ($F_{4,184}$ = 33.89, $ p <$ 0.001, $\eta^2$ = 0.42). Furthermore, post-hoc comparisons indicate that when conditions are grouped by $v$, all $D$ pairs are significantly different (all $t_{184}$ $>$ 7.874,  $p <$ 0.001) with a decrease of $T_{\rm dec}$ for larger values of diffusivity (see Appendix~\ref{sec:lin_mix_eff} for details).

\section{Thermodynamic trade-offs}
\label{sec:TUR}
\begin{figure*}
\includegraphics[width=0.95\textwidth]{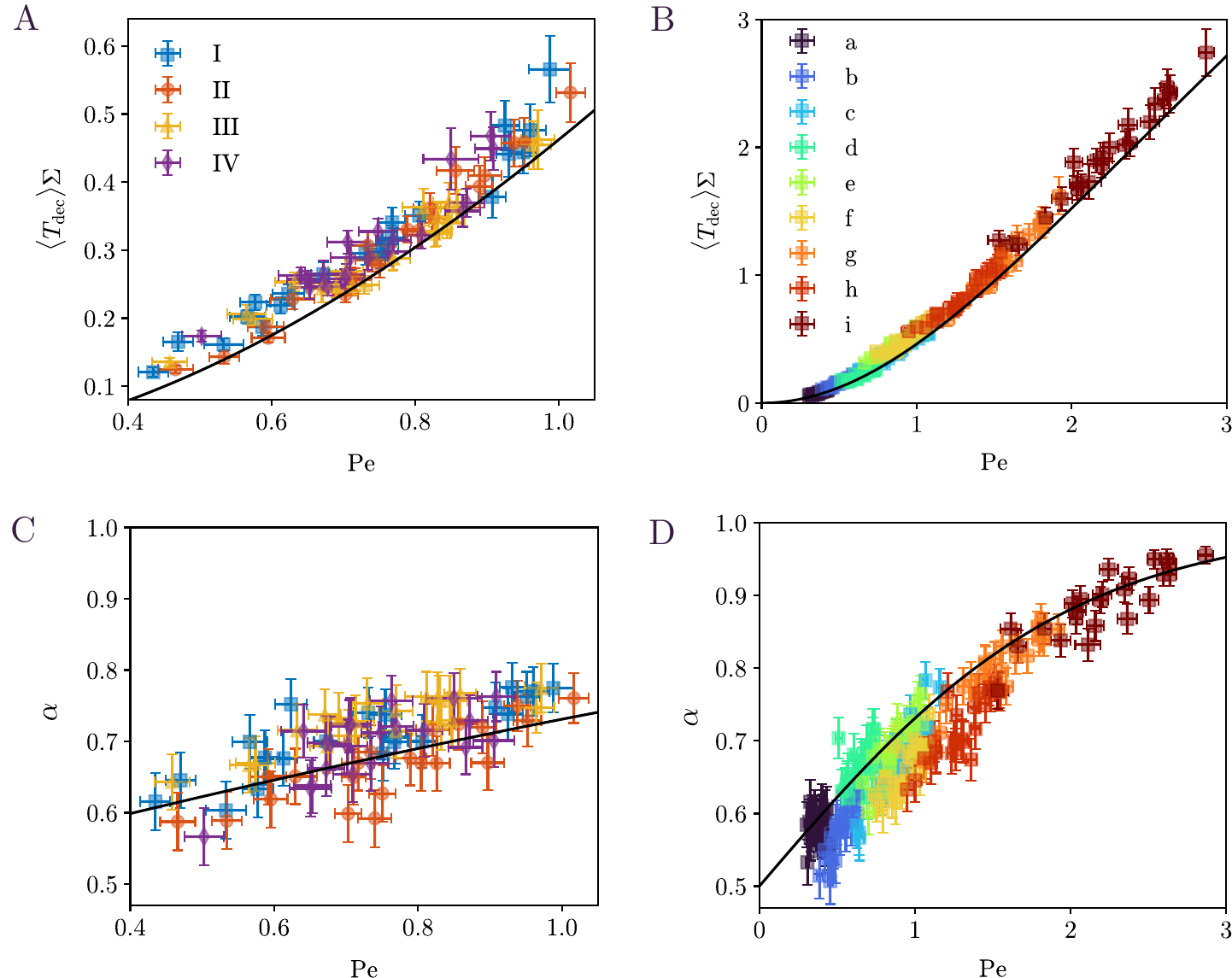}
\caption{{\bf Experimental test of the stochastic-thermodynamics relations Eq.~\eqref{eq:TURdt} and~\eqref{eq:acto} in human perceptual decision making.} (A, B) Mean decision time (scaled by the rate of dissipation $\Sigma=v^2/D$)  as a function of the participants' effective P\'eclet number for Experiment A (A), and Experiment B (B). The black line is given by $\rm{Pe} \tanh{(\rm{Pe}/2)}$ [Eq.~\eqref{eq:TURdt}]. (C, D) Mean accuracy as a function of the participants' effective P\'eclet number for Experiment A (C), and Experiment B (D). The black line is given by the right-hand side in Eq.~\eqref{eq:acto}. (A-D) Each  symbol is obtained from the statistics of a single participant, and different shapes and colors are used to denote different conditions. The effective P\'eclet number for each participant is estimated from the mean absolute value of the participant's decision threshold in each condition as $\text{Pe}  =v \langle |X_{\rm dec}|\rangle /D$. The error bars correspond to SEM. }
\label{TUR_SPRT}
\end{figure*}

\begin{figure}
\includegraphics[scale=0.7]{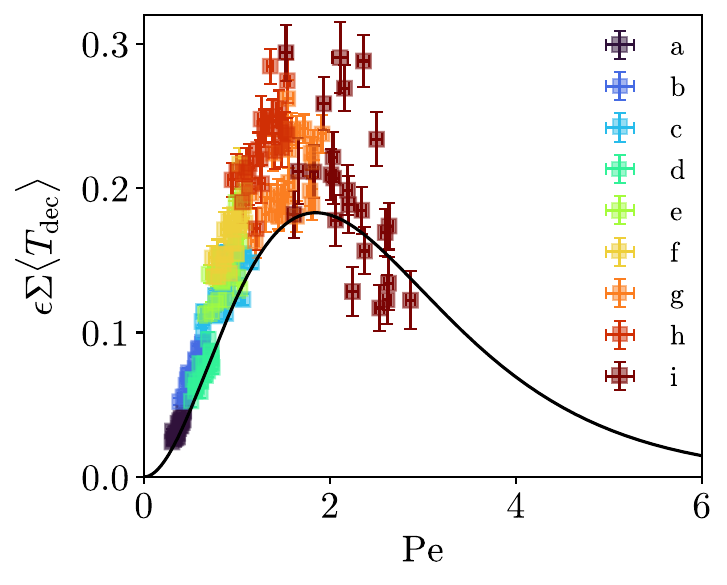}
\caption{{\bf Experimental test of Eq.~\eqref{eq:TUR_error} in human perceptual decision making.} The product of the mean decision time, rate of dissipation and error probability as a function of the participants' effective P\'eclet number for Experiment B in different conditions (see legend). The black line is given by the right-hand side of Eq.~\eqref{eq:TUR_error}, which reaches a maximum at~$\text{Pe}\simeq 1.85$. Error bars are SEMs. }
\label{TUR_SPRT_all}
\end{figure}

We verify SPRT as the upper limit of human performance by inspecting the statistics of each participant under each of the conditions tested in both Experiments A and B. Figs.~\ref{TUR_SPRT}A and~\ref{TUR_SPRT}B show that the mean decision time of each participant in each condition satisfies the thermodynamic uncertainty relation conjectured by Eq.~\eqref{eq:TURdt} for both experiments. Here, the fundamental lower bound of the right-hand side of Eq.~\eqref{eq:TURdt} is established using an effective P\'eclet number  $\text{Pe}  =v \langle |X_{\rm dec}|\rangle /D$ associated with the statistics of each participant in each trial.
  Moreover, in  Figs.~\ref{TUR_SPRT}C and ~\ref{TUR_SPRT}D we show that the accuracy of each participant in all conditions is well characterized by Eq.~\eqref{eq:SPRT_accuracy}. In particular, in Fig.~\ref{TUR_SPRT}D participants' accuracy tends to follow an upper bound given by the SPRT accuracy associated with the effective Peclet number,
  \begin{equation}
    \alpha \leq \frac{1}{ 1+ e^{-\mathrm{Pe}}}.
    \label{eq:acto}
  \end{equation}
Thus, the results suggest that the SPRT  constrains not only decision times, but also accuracy/error probability. Combining Eqs.~\eqref{eq:TURdt}  and \eqref{eq:acto}, a trade-off relation between error probability, dissipation, and decision time can be conjectured,
\begin{equation}
   \epsilon \Sigma \langle T_{\mathrm{dec}}   \rangle \geq \mathrm{Pe} \frac{\left(e^{\mathrm{Pe}}-1\right)}{\left(e^{\mathrm{Pe}}+1\right)^2},
   \label{eq:TUR_error}
\end{equation}
where $\epsilon = 1 - \alpha$ is the error probability. In Fig.~\ref{TUR_SPRT_all}, we find that theoutcomes of Experiment B fulfill Eq.~\eqref{eq:TUR_error}  for nearly all experimental conditions. 
This trade-off is influenced by the transport properties of the stimulus through Pe, with the right-hand side reaching a maximum at an ``optimal'' $\text{Pe}\simeq 1.85$ at which the dissipation needed to attain a given error probability and decision time is maximized. The few data points that deviate most from relation~\eqref{eq:TUR_error} correspond to high velocity $v$ and low diffusivity $D$, conditions for which the accuracy of the participants is nearly $\sim 90\%$, leading to unreliable statistics concerning errors.

In the data presented so far 
decision times are longer than those predicted by SPRT given the achieved accuracy level, betraying suboptimal decision making. Additionally, trial-by-trial decision thresholds span a wide distribution of thresholds across experimental conditions (e.g., see histograms of decision thresholds in Figs.~\ref{Fig1}C--F for one participant). The variance of such a distribution is significantly larger than that of a control experiment where thresholds are displayed as vertical lines on the screen throughout the movies; see Appendix~\ref{sec:var_in_decision_threshold}.
This analysis discounts the possibility fluctuations in the decision threshold can be solely attributed to variability in mere reaction time; instead fluctuations must be rooted in the participants' interpretation of  complex sensory evidence.

SPRT assumes fixed decision thresholds. Alternative approaches have included noise in the starting point of the likelihood ratio as an additional parameter to the SPRT~\cite{ratcliff2008diffusion}. However, in our experiments, the starting position of the disk is refreshed anew to $X_0=0$, $Y_0=0$ at each trial. The need to explain suboptimality motivates us to go one step beyond traditional approaches and implement an evidence integration model.

\section{Evidence Integration Model and memory inference}
\label{sec:EIM}

\begin{figure}[tbp]
\centering\includegraphics[scale=0.5]{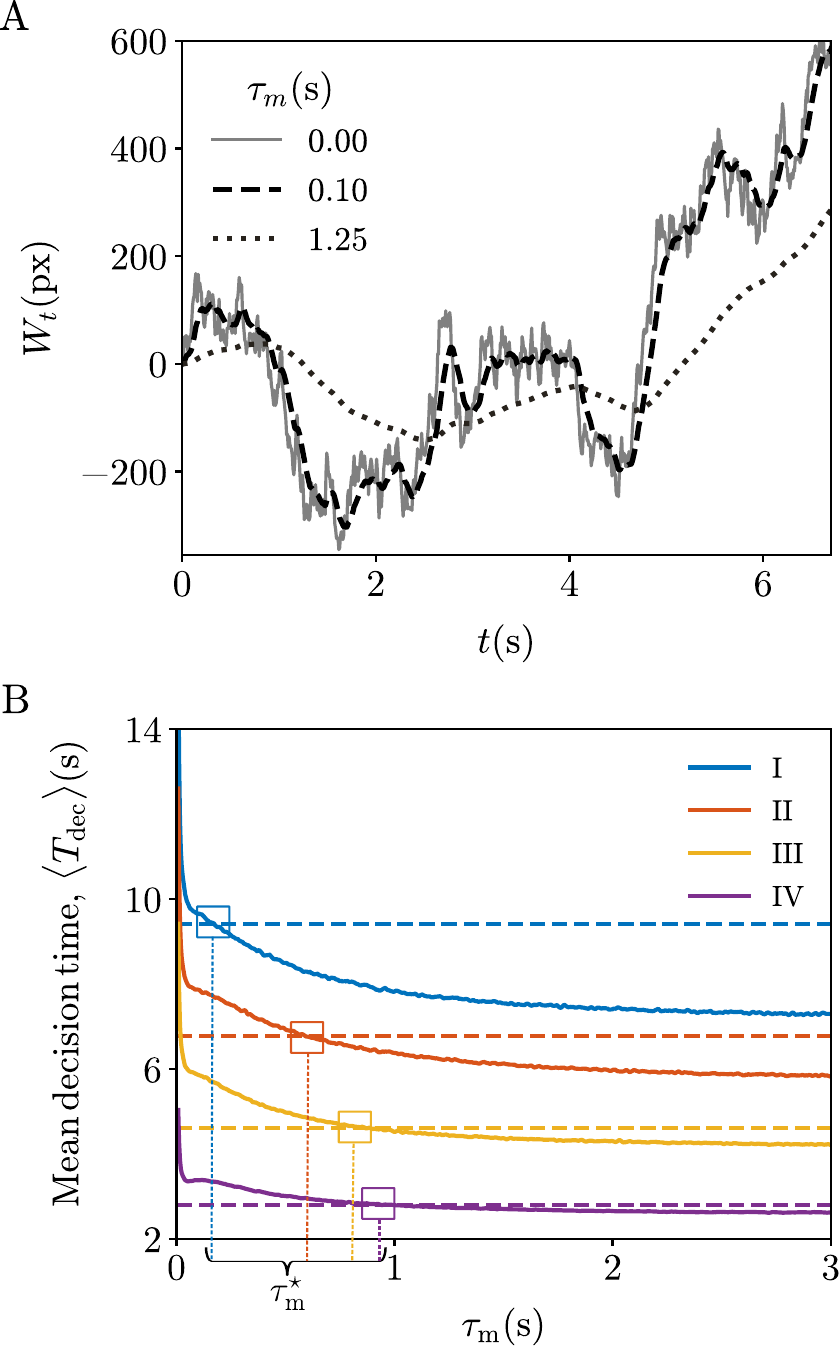}

\caption{ {\bf Evidence Integration Model (EIM).} (A) Sample trajectories for the decision accumulator $W_t$ associated with a given trajectory of the position $X_t$ for different memory relaxation times $\tau_m$. Note that for $\tau_m = 0$, $W_t=X_t$ (gray solid line). (B) Inference of the optimal memory relaxation time $\tau_m^\star$  for one participant in the four conditions:  experimental mean decision time (horizontal dashed lines), and mean escape time of $W_t$ from the interval $(-L_W,L_W)$ extracted from numerical simulations (solid lines) as a function of $\tau_m$. Here $L_W=\langle\vert W_{T_{\rm dec}}\vert \rangle$ is determined from experimental trajectories $X_{[0,T_{\rm dec}]}$ (see Fig.~\ref{Fig1}). For each condition and each participant, $\tau_m^\star$ is the value of $\tau_m$ yielding the minimal difference between the experimental and simulated mean decision times, visualized as the crossing point within the colored boxes. 
}
\label{fig4}
\end{figure}

One possible explanation is limited memory. SPRT assumes that participants accumulate $X_t$ as evidence without loss or ``leakage" and thus does not account for the memory dynamics that might constrain how the brain integrates evidence. To embody possible memory limits, we introduce an Evidence Integration Model (EIM) comprised of an additional parameter reflecting memory limits. In the EIM, the collection of evidence is given by an accumulator $W_t$ that is a function of $X_{[0,t]}$ defined by 
\begin{align}
    W_t \equiv \frac{1}{\tau_m} \int_{0}^{t} {\mathrm{d}s} \, \mathrm{e}^{-s/\tau_m} \, X_{t-s} \, , \label{eq:Wtdef_backward}
\end{align}
where $\tau_m$ is the {\em memory relaxation time}. Thus, the accumulated evidence,  $W_t$, is a weighted moving average (an ``interpolation") of a set of previous values of the position. The exponential weighting defined by $\tau_m$ has the effect of assigning greater impact at any given time to the most recent observations of position. In other words,
the total quantity of information in  Eq.~\eqref{eq:Wtdef_backward} is read backward in time, giving exponentially less weight $\mathrm{e}^{-s/\tau_m}$ to the past evidence $X_{t-s}$ with respect to the evidence at time $t$. 
From Eq.~\eqref{eq:Wtdef_backward} one may write $\tau_{m} \dot{W}_{t} = - W_t + X_t$. For the stimulus $X_t$ given by the stochastic process $\dot{X}_t=Hv+\sqrt{2D}\xi_t$, the accumulator $W_t$ obeys an underdamped Langevin equation
\begin{equation}
       \tau_m \ddot{W}_t = -\dot{W}_t +Hv+\sqrt{2D}\xi_t,
       \label{eq:ud}
\end{equation}
in which $\tau_m$ acts as an effective inertial relaxation time. Fig.~\ref{fig4}A shows trajectories of the accumulator~$W_t$ for  one instance of the stochastic stimulus $X_t$ for different values of $\tau_m$. The short memory limit $\tau_m\to 0$ corresponds to the overdamped limit in Eq.~\eqref{eq:ud}, in which $W_t \to X_t$, while for larger $\tau_m$ underdamped dynamics are generated, making the accumulator a smoothed version of the stimulus. To fully characterize the model, we establish the {\em EIM decision criterion} as the first exit of $W_t$ through either boundary of a symmetric interval $(-L_W,L_W)$: decision $D_{\rm dec} =+1$ occurs if $W_t$ reaches $L_W$ before $-L_W$, and vice versa for decision $D_{\rm dec} = -1$. This results in a decision-making model with two free parameters, $\tau_m$ and $L_W$, both to be fitted from the experimental data. 

Following canonical approaches in cognitive neuroscience, we expect both decision threshold $L_W$ and memory relaxation time $\tau_m$ to be heterogeneous across participants and experimental conditions. As we will show, $L_W$ and $\tau_m$ must be participant-specific in order to account for the inter-participant and inter-condition statistics.  Therefore, to fit the EIM to the data, we infer participant-specific values of the memory relaxation time $\tau_m^\star$ and the decision threshold $L_W^{\star}$ for each experimental condition.
To infer $\tau_m^\star$ for a participant in a given condition, we explore a range of possible values of $\tau_m$ (see Fig.~\ref{fig4}B) determined empirically. For each value of $\tau_m$, evaluating Eq.~\eqref{eq:Wtdef_backward} along the trajectories $X_{[0, T_{\mathrm {dec}}]}$ extracted from the experiment, we obtain the decision threshold $L_W=\langle \vert W_{\rm dec}\vert\rangle=\langle \vert W_{T_{\rm dec}} \vert\rangle$ as the average of the absolute value of the decision accumulator at the decision time. 
From this, we have a collection of parameter pairs $\{(\tau_m, L_W)\}$.
For each pair $(\tau_m, L_W)$, we employ the EIM criterion to obtain from numerical simulations the mean exit time of $W_t$ from $(-L_W, L_W)$. 
As illustrated for the data fitted from an individual participant in Fig.~\ref{fig4}B, the intrinsic memory time $\tau_m^\star$ is identified as the mean exit time obtained from the simulations (solid traces) yields the closest value to the experimentally observed mean decision time (horizontal dashed lines in  Fig.~\ref{fig4}B). The locations of the closest values are highlighted by the colored box on each trace. The corresponding value of $L_W$ is chosen as $L_W^{\star}$. See Appendix~\ref{EIM_sup} for more details on the fitting procedure.

\begin{figure}%
\centering
\includegraphics[width=0.5\textwidth]{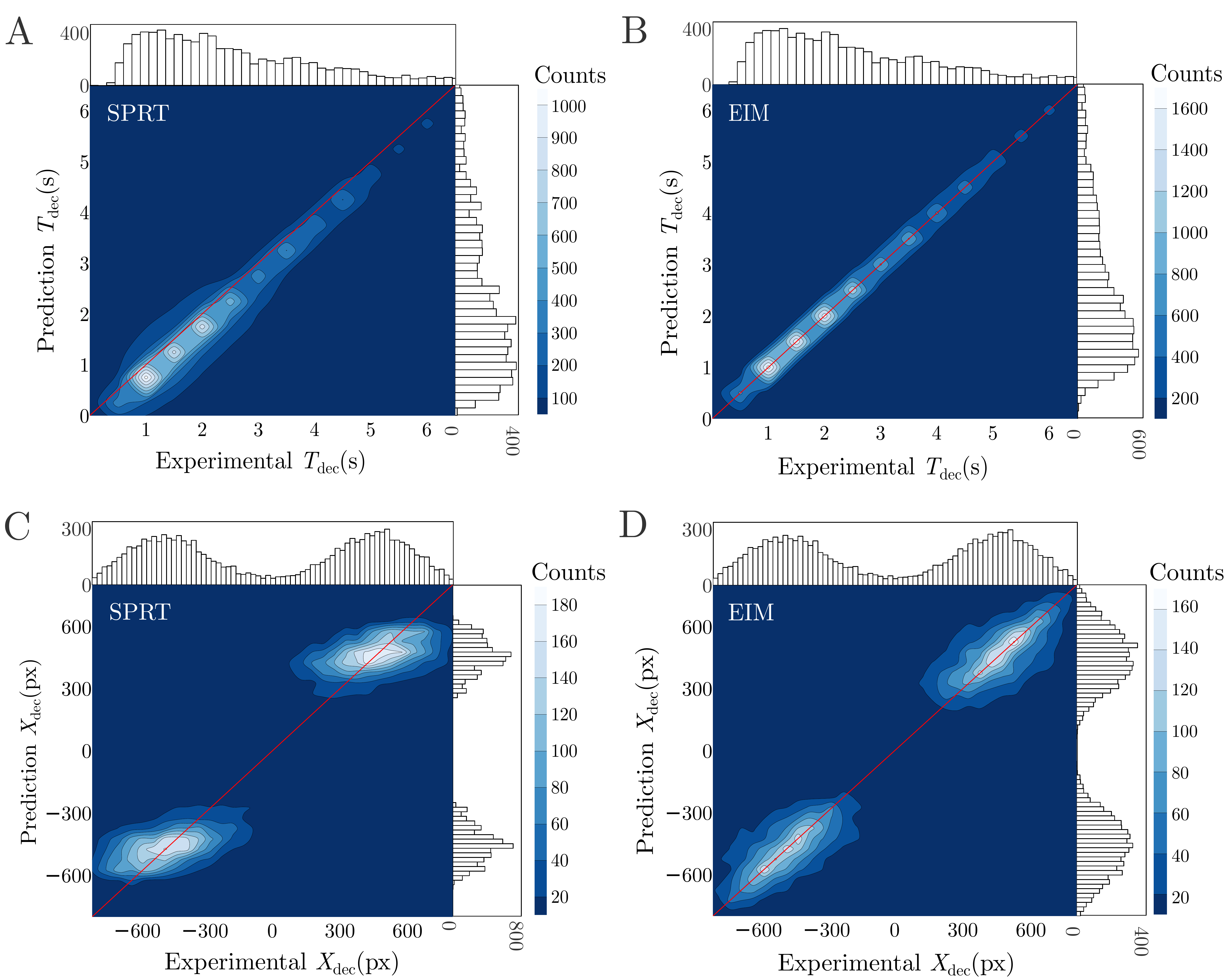}
\caption{{\bf Comparison of decision making models to experimental data.} 
(A-B) Joint histograms of  the trial-by-trial observed decision time $T_{\rm dec}$  for all participants and conditions, and the numerical predictions of $T_{\rm dec}$ by SPRT (A) and by EIM (B). (C-D) Same format as (A-B) but for observed decision threshold $X_{\rm dec}$. The red lines have slope one, representing perfect agreement between the experimental output and model predictions.}
\label{fig5}
\end{figure}

We validate the EIM by comparing its predicted decisions with the experimental data in trial-by-trial resolution (see Fig.~\ref{fig5} for Experiment A and Fig.~\ref{color_map_exp2} in the Appendix for Experiment B). Figs.~\ref{fig5}A--B show joint histograms of predicted and observed decision times collected from all trials of all participants, under all conditions, for the SPRT (Fig.~\ref{fig5}A) and EIM (Fig.~\ref{fig5}B). See Appendix~\ref{sec:cross_validation} for additional analysis related to the robustness of decision thresholds used to predict decision times. We proceed similarly for the joint histograms of decision thresholds in Figs.~\ref{fig5}C--D comparing experimental outputs with predictions from the SPRT (Fig.~\ref{fig5}C) and EIM (Fig.~\ref{fig5}D). Note that the distribution of $X_{\rm dec}$ associated with the SPRT shows substantial variation -- its horizontal spread -- despite the fact that such decisions are taken when $X_t$ crosses fixed thresholds. See Appendix~\ref{sec:var_in_decision_threshold} for further analysis of $X_{\rm dec}$ variance. The Kolmogorov-Smirnov (KS) test is used to compare the distribution of decision times and thresholds predicted by the SPRT and the EIM against the experimental distributions. For decision times, the KS test reveals significant deviations between SPRT predictions and experimental data ($D_{11963}$ = 0.186, $p =$ 0.012), while the same test uncovers no significant difference between the EIM predictions and experimental data ($D_{11963}$ = 0.103, $p =$ 0.421). The SPRT difference arises from its systematic trend of predicting faster decisions (below the diagonal) compared to the experimental observations (Fig.~\ref{fig5}A). For decision thresholds, EIM predictions again outperform SPRT ($D_{11963}$ = 0.118, $p =$ 0.273 vs. $D_{11963}$ = 0.262, $p <$ 0.0001, respectively). This analysis demonstrates that the EIM captures inter-trial variability more effectively than SPRT. 

We also perform an analysis with an alternative model for evidence integration, the Ornstein-Uhlenbeck (OU) model. The OU model that we develop replaces the  position $X_{t-s}$ by the velocity $\dot{X}_{t-s}$ in the kernel of Eq.~\eqref{eq:Wtdef_backward}, i.e. $W_t \equiv \frac{1}{\tau_o} \int_{0}^{t} {\mathrm{d}s} \, \mathrm{e}^{-s/\tau_o} \, \dot{X}_{t-s}$ which leads to an overdamped Langevin equation for the accumulator
\begin{equation}
       \tau_o \dot{W}_t = -W_t +Hv+\sqrt{2D}\xi_t,
       \label{eq:OU_pro2}
\end{equation}
whose solution is a Markovian stochastic process with relaxation time $\tau_o$. 
Appendix \ref{sec:OU_model}  shows that it is not possible to find values for  $\tau_o$ that fit the experimental data in most of the experimental conditions for which the EIM, by contrast, provides accurate fits and predictions. This highlights the fact that non-Markovian features in evidence accumulation are essential to capture the observed decisions.

\begin{figure}
\centering
\includegraphics[scale=0.5]{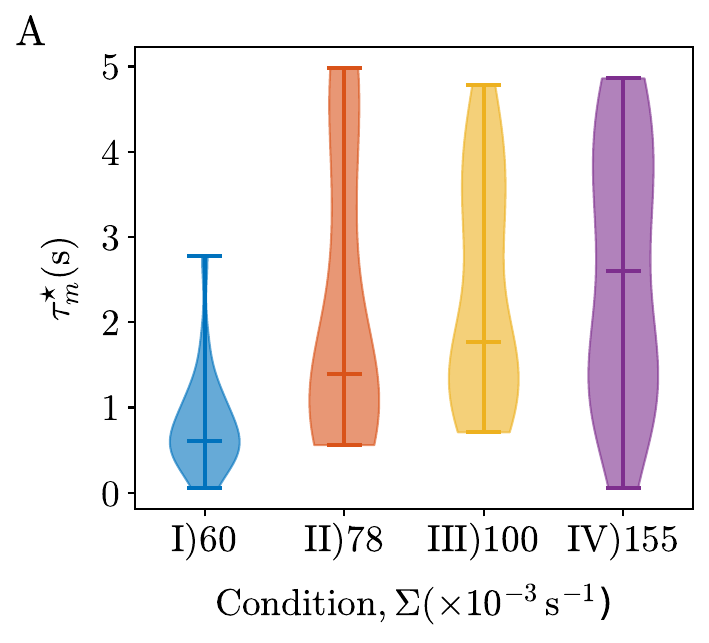}
\hspace{0.25cm}
\includegraphics[scale=0.5]{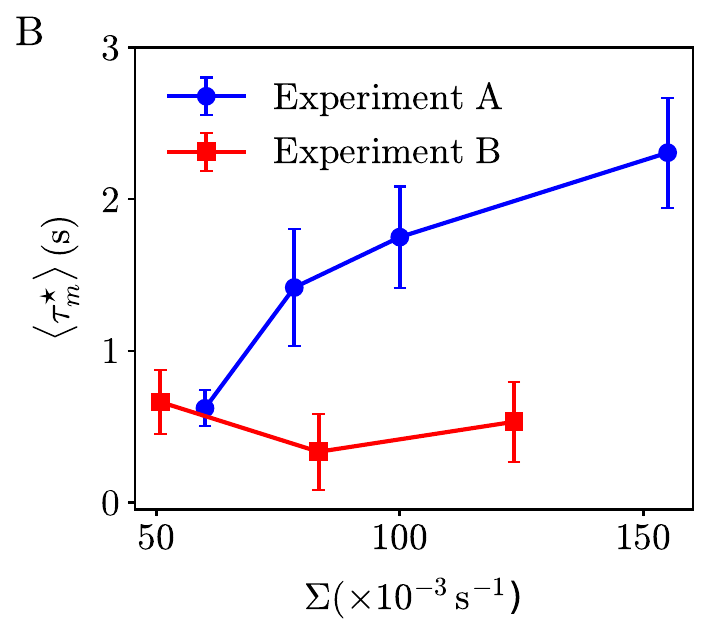}
\caption{ {\bf Intrinsic memory relaxation time  $\tau_m^{\star}$ inferred from fitting the EIM to human participants' decisions.} (A) Distribution of the inferred $\tau_m^\star$ for all participants in each condition of Experiment A. The central horizontal lines are set at their respective medians. 
(B) Comparison of the inferred value of $\tau_m^{\star}$  as a function of the conditions of Experiment A (blue), and three corresponding conditions from Experiment B (red), with error bars given by SEM across participants. 
}
\label{tau_ms_compar}
\end{figure}

 What is the functional meaning of $\tau_m^\star$ , the parameter that causes the EIM to capture inter-trial variability more effectively than the SPRT?  Fig.~\ref{tau_ms_compar}A gives the distributions of $\tau_m^\star$ for each participant in Experiment A. Interestingly, the median of $\tau_m^\star$ over all participants increases with the entropy production rate $\Sigma$, indicating that participants deploy a memory trace with longer time constant for trajectories with a higher degree of nonequilibrium. This strategy aligns with the need for more cautious decision-making under higher entropy production: longer memory relaxation times allow participants to accumulate evidence over a larger window.

In contrast, the diagonal conditions of Experiment B do not reveal the same relationship of increasing $\tau_m^\star$ with increasing $\Sigma$ (Fig.~\ref{tau_ms_compar}B). These results suggest that in the block design of Experiment A the optimal memory relaxation time may be adapted to the $\Sigma$ of the ongoing set of 50 trials, allowing the $\Sigma$-specific $\tau_m^\star$ to be carried forward across trials. In other words, the participant initiates each trial in Experiment A with a predicted stimulus $\Sigma$  and a corresponding internal $\tau_m^\star$. However, the intermixed design of Experiment B prevents the participant from initiating trials with optimally ``preset" $\tau_m^\star$. The setting of the memory relaxation time appears not to be triggered ``bottom-up" from the statistics of the stimulus in real time. Calibrating $\tau_m^\star$ may require more time than is available within a single trial. We speculate that the participants in the intermixed design therefore may act with a short $\tau_m^\star$, an effective compromise to allow acceptable accuracy for any encountered $\Sigma$.

\section{Discussion}
\label{discussion}

On the basis of the statistics obtained from human participants in a perceptual decision making task, this work shows that judgments of nonequilibrium phenomena are constrained by fundamental thermodynamic trade-offs between decision time, accuracy and entropy production. The study constitutes the first exploration in cognitive neuroscience of the recently theorized thermodynamic uncertainty relations for Markovian systems~\cite{Barato2015,roldan2015decision,Falasco2020,Horowitz2020}, and thus introduces the potential utility of such theories for exploring the thermodynamics of non-Markovian nonequilibrium phenomena.

Human perceptual decision making data are typically investigated by fitting the accumulation of an unobservable variable, known as evidence, by parameters like drift rate and choice threshold. Our study, on the other hand, adopts an alternative approach wherein the evidence variable (the stimulus position) is directly observable. Having in hand the actual quantity of available evidence -- a quantity that is not concretely defined in motion coherence decision making studies 
-- allows us to apply stochastic thermodynamics principles to observed data to quantify the deviation of decisions from SPRT's optimality. Measured deviations from optimality of decision-times and decision-thresholds are small, yet sufficient to justify the search for a model to explain the brain's suboptimality.

Because the nervous system, unlike SPRT, is characterized by non-optimal information storage~\cite{dorpinghaus2023optimal}, we built a ``forgetting'' function, referred to as EIM, into the decision model. EIM, once its memory parameter is tuned, shows accurate prediction of participants' choices on a trial-by-trial basis. In particular, the fact that alternative Markovian models (OU) fail to describe our data demonstrates the non-Markovian nature of evidence accumulation in our experimental assays. Moreover, the $\Sigma$-dependent time constant suggests that participants adjust their memory relaxation time to the dissipation rate of the observed phenomenon: in Experiment A participants establish a memory window with a longer time scale for stimuli that are far from equilibrium. Adjusting the time scale of evidence integration has been suggested as a strategy to overcome the volatility of the signal by averaging out noise to avoid false alarms~\cite{ossmy2013timescale, glaze2015normative, piet2018rats}. It is tempting to speculate that the time constant governing the ``leakage" of sensory evidence in the current study may be related to the time constants identified in earlier work, where participants' estimates of stimulus intensity or duration are modeled as the leaky accumulation of noisy sensory evidence~\cite{diamond2023tactile, toso2021sensory, hachen2021dynamics, fassihi2020making, zuo2019texture, esmaeili2019neuronal, fassihi2017transformation}.

It is worthwhile to compare our findings with other studies on optimality in decision making. Recent work has revealed how the deviation of human decision-making from suboptimality may be reduced by over-training ~\cite{balci2011acquisition,bogacz2006physics,bogacz2010humans,starns2010effects,zacksenhouse2010robust}.   In the present study, participants are not over-trained, nor are they  verbally informed about the experimental manipulations. This means they can access the parameters of the stimuli across conditions only by implicit computations. Second, the trial numbers are fixed in a session, and the participants are not rewarded for faster decisions. This is different from studies with fixed session time, where participants receive more rewards for undertaking more trials, potentially overweighting speed. \newline \newline
We can thus summarize the main findings of our work as follows. (1) In a new moving disk paradigm, the quantity of available evidence is explicit through principles of stochastic thermodynamics. This paradigm, differently from the classical random dot motion experiment, allows the true ideal observer performance to be computed.
(2) In Experiment A, ideal performance quantified by SPRT exceeds observed human performance, betraying a small but systematic suboptimality. (3) Observed suboptimality is mimicked by including memory constraints in a new model termed EIM, (4) Participants adjust $\tau_m^\star$ (memory relaxation time parameter of the EIM) based on the entropy production rate of the stimuli. (4) In Experiment B we find that intermixed presentation of stimuli with different entropy production rates diminishes performance by inducing participants to apply a single memory time constant $\tau_m^\star$ across trials; the time constant apparently is not reset within each ongoing trial.

Exploring the neural dynamics governing perceptual decision-making during non-equilibrium fluctuations offers a promising venue for future research. For instance, NMDA-mediated synapses have been highlighted as potential modulators of integration time windows in the brain and a future work modeling the memory relaxation time across conditions and experiments might implement these for a biologically plausible mechanism~\cite{wong2008temporal,wang2002probabilistic, wong2006recurrent}. In addition to the neurobiological mechanisms, the dynamics of macroscale brain activity for perceptual decision making of non-equilibrium fluctuations is also open for future inquiry. The most robust insights to date have revealed, using brain imaging methods, that the dynamics of the human and primate brain exhibits speed-accuracy trade-offs~\cite{fang2019nonequilibrium} and a measurable degree of time irreversibility~\cite{lynn2021broken} that depends on the type of cognitive tasks~\cite{lynn2021broken}, consciousness levels~\cite{g2023lack,de2023temporal}, and pathology in the brain~\cite{cruzat2023temporal,zanin2020time}. These perspectives are promising not only for advancing our understanding of human cognition but also for lighting the broader principles that govern complex systems, taking swift and accurate decisions in a changing world.

\section*{Acknowledgements}
MED acknowledges financial support from Human Frontier Science Program, project RGP0017/2021, European Union Horizon 2020 MSCA Programme NeuTouch under Grant Agreement No 813713, Italian Ministry PRIN 2022 Contract 20224FWF2J. DB  acknowledges financial support from the European Research Council--ERC (Grant Agreement No 682117, BiT-ERC-2015-CoG), the Italian Ministry of University and Research under the call FARE (project ID: R16X32NALR) and under the call PRIN2022 (project ID: CCPJ3J). ER acknowledges financial support from PNRR MUR project PE0000023-NQSTI. We thank Antonio Celani, Stefano Ruffo, and Jin Wang for valuable discussions, and Marta Fedele for her help in the pilot phase of the study.

\section*{Author Contributions}
A.D. conceptualization, data curation, investigation, writing-original draft; A.D. and Y.S. formal analysis, software, validation, visualization, methodology, writing-review and editing; G.F. conceptualization, formal analysis, methodology, software, validation, supervision, writing-review and editing; D.D. formal analysis, methodology, validation, supervision, writing-review and editing; M.E.D. conceptualization, formal analysis, methodology, supervision, writing-review and editing; D.B conceptualization, methodology, funding acquisition, project administration, resources, supervision, writing-review and editing; E.R. conceptualization, formal analysis, validation, methodology, funding acquisition, project administration, resources, supervision, writing-original draft, writing-review and editing. All authors approve the final manuscript. 

\newpage\newpage
\section*{APPENDIX}
\appendix
\renewcommand{\thefigure}{\arabic{figure}}

\section{Materials and Methods}
\label{materials}

\textbf{Experiment A.} The experiment comprises four experimental conditions, each characterized by prescribed values of drift velocity and diffusivity (see Table~\ref{tab:1}). 
\begin{table}[!htbp]
\centering
\begin{tabular}{ |c|c|c|c| } 
\hline
Condition & $v\, (\text{px}/\text{s})$ & $D\, (\text{px}^2/\text{s})$ & $\Sigma\, (\text{s}^{-1})$\\
\hline
I & $36$ & $21600 $ & $0.060$ \\ 
II & $47$ & $28200$ & $0.078$ \\ 
III & $60$ & $36000$ & $0.100$\\ 
IV & $93$ & $55800$ & $0.155$\\ 
\hline 
\end{tabular}
\caption{Experiment A, parameter values for the drift velocity $v$, diffusivity~$D$, and entropy production rate $\Sigma=v^2/D$ used in the experimental conditions (I, II, III, and IV) .}
\label{tab:1}
\end{table}
The parameter values are chosen such that across conditions, the entropy production rate $v^2/D$ varies, yet the ratio $v/D\simeq 1.67\times 10^{-3}$ px$^{-1}$ remains constant. 
 For each condition and each participant, we carry out $3$ blocks of $50$ trials each, summing up to 12 blocks of 50 trials, that is, $600$ trials in total. 
All participants are displayed with the same set of $600$ trial movies. However, the order of the blocks and the trials within a block are randomized for each participant.  
Participants are asked to sit in a room with dim light in front of the lab computer equipped with a 2233RZ series business monitor with a spatial resolution of [1680 px $\times$ 1050 px] and a refresh rate of $120$ Hz.  Throughout the experiments, a chin rest is used to restrict head movements and fix head position at a distance of $57$ cm from the computer screen. 

The stimuli (a disk with a diameter of 2° degrees of visual angle) are located in the center of the screen at the beginning of the trials.
Each disk trajectory is generated by performing a Euler numerical integration of the Langevin equation $\dot{X}_t=Hv+\sqrt{2D}\xi_t$ and $\dot{Y}_t = \sqrt{2D}\eta_t$ with $H=\pm 1$, a binary random variable sampled with equal probability.
We implement reflecting boundary conditions along the $y$-axis of the screen. In other words, the disk bounces back to its previous position when it touches any of its edges along the $y$-axis.  
The discrete time step in numerical simulations is chosen to be the inverse of the screen temporal resolution.

\begin{figure*}
\centering
\includegraphics[scale=0.45] {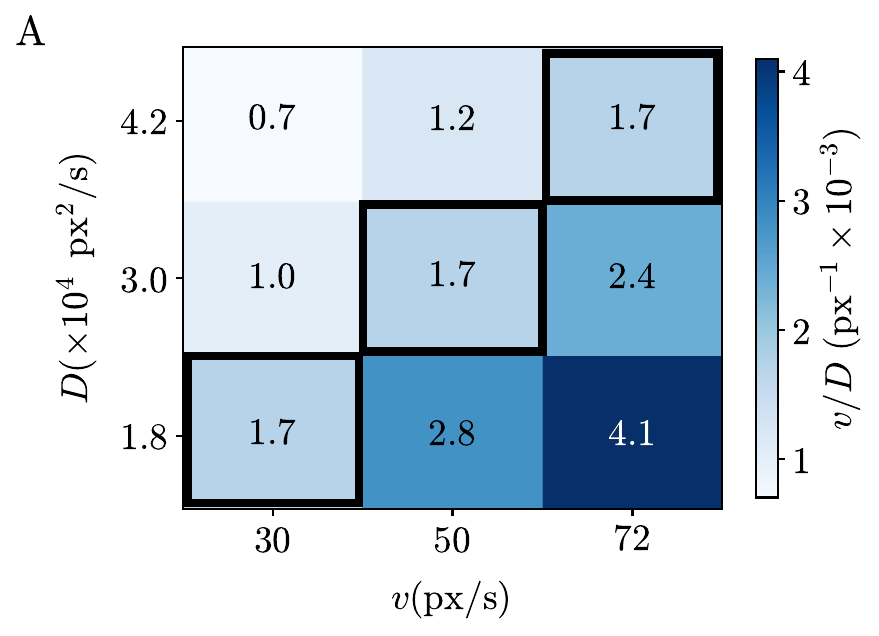}
\hspace{0.25cm}
\includegraphics[scale=0.45]{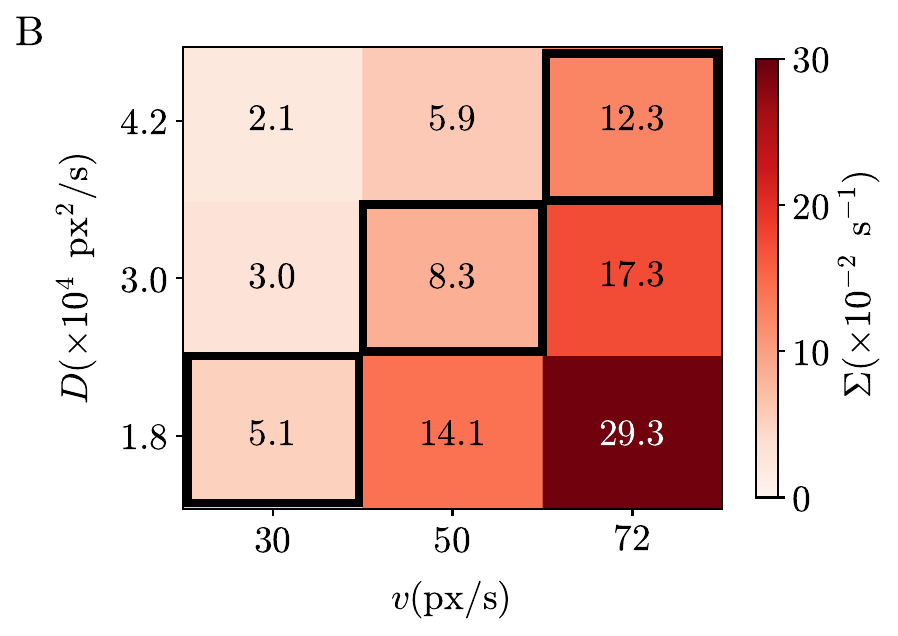}
\caption{ {\bf Parameter values for conditions in Experiment B.} (A) Parameter values for drift velocity $v$, diffusivity $D$, and their ratio used in the nine experimental conditions in Experiment B. Each condition is a combination of the $v$ and $D$ of the corresponding column and row of the table, shown with its $v/D$ value color-coded with its numerical value shown in text. The diagonal conditions with the black frame refer to the conditions with a fixed value $v/D$, and these three conditions have parameters similar to that of Experiment A.  (B) Same experimental conditions but numerical values and color-code show the corresponding entropy production rate $\Sigma$ for each experimental condition.}
\label{FigPrmSpcTable}
\end{figure*}

\textbf{Experiment B.} This experiment consists of nine experimental conditions, each characterized by a combination of three velocity and three diffusivity values (see Fig.~\ref{FigPrmSpcTable}). These parameter values are chosen such that across the three diagonal conditions (shown in black frames in Fig.~\ref{FigPrmSpcTable}), $v$ and $D$ vary with a constant ratio of $v/D\simeq 1.69\times 10^{-3}$ px$^{-1}$. These diagonal conditions are designed to compare the two experiments.
For each condition and each participant, we carry out $144$ trials each, presented as 36 blocks of 36 trials, that is, $1296$ trials in total. 

\par
All participants are shown with the same set of trial movies. However, the order of the blocks and the trials within a block are randomized for each participant. In addition (unlike Experiment A), in this experiment, trials are presented in an intermixed block design, meaning that in each block, participants observe trials from all nine conditions in random order. In each block, there are a fixed number of trials (four) of the movies from each condition. This is done to avoid any condition dominating others in a block. \par
The lab computer is equipped with a BenQ ZOWIE XL2731 monitor with a spatial resolution of [1920 px $\times$ 1080 px] and a refresh rate of 144 Hz. \par
Everything else not indicated here is the same as in Experiment A.

\par
\textbf{Data analysis.} \textit{Experiments A and B.} All data analyses are performed in~R, Python, and MATLAB. The missed trials correspond to the cases where the participant has not taken a decision before the disk reaches the screen edge from either side (right/left). Prior to the analyses, we remove these missed trials ($\sim 5\%$ for Experiment A and $\sim 3\%$ for Experiment B) from the dataset.  We use the lmer program from the lme4 package~\cite{Bates2015-ts} for linear mixed-effects modeling. In Experiment B, $\sim 0.06\%$ of the outlier trials ($T_{\rm dec}$ $<$ 150 ms) are removed prior to the linear mixed-effects modeling; there are no such trials in Experiment A.

\section{Experimental details}
\label{sec:addexp}

\textbf{Participants.} All the participants are tested in the laboratory of the International School for Advanced Studies, SISSA, Trieste. Prior to the experiments, the written consent form approved by the Ethics Committee of the SISSA is given and obtained from the participants. At the end of the experiments, all participants are provided with a debriefing session. \par
\textit{Experiment A.} Twenty one right-handed participants, aged between $23$ and $59$, with normal or corrected-to-normal vision are recruited ($N_{\rm Female} = 12$, ${\rm M} = 30.76$, ${\rm SD} = 8.8$) on a voluntary basis for this experiment. The participants are tested for approximately 1 to 1.5 hours depending on their own pace. \par
  \textit{Experiment B.} Twenty four participants (1 left-handed), who have not participated in Experiment A, aged between $19$ and $35$, with normal or corrected-to-normal vision are recruited ($N_{\rm Female} = 11$, ${\rm M} = 27.17$, ${\rm SD} = 3.48$) for Experiment B. The participants are tested for approximately 4 hours separated into 2 sessions conducted on different days. In each session, participants complete half of the blocks. Participation is compensated by 10 euro per hour, upon completion of the experiment.

\begin{figure}[!htbp]
\centering
\includegraphics[scale=0.3]{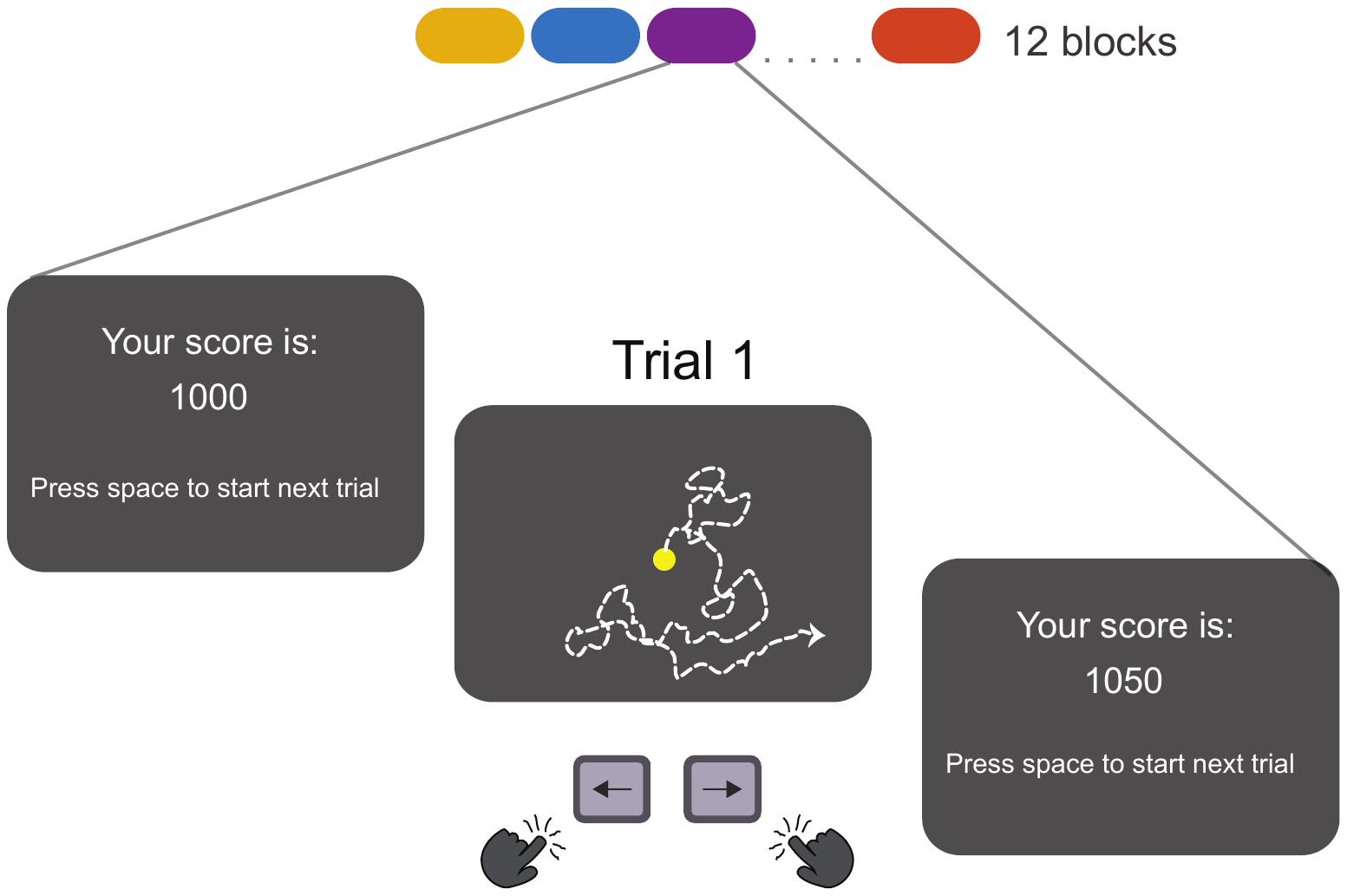}
\caption{\textbf{Schematics of visual displays in Experiment A}. Each of the twelve blocks consists of trials from the same condition. For each condition, the trials are shown in three blocks. In the beginning of each block, participants start with 1000 points. The score is accumulated within each block as participants give correct, incorrect, or missed responses, and in the next block participants start with 1000 points afresh.}\label{S3}
\end{figure}

\textbf{Procedure and data collection.}
Prior to Experiments A and B, participants undergo a training session to familiarize themselves with the task. 
The experiment is designed as a two-alternative forced choice task. As soon as the trial starts, participants are required to report the direction of the disk by pressing the right (left) arrow key corresponding to the right (left) direction before the disk reaches any of the edges of the screen. After each trial, participants receive immediate feedback on any of the three possible outcomes they can get from their response together with an accumulating score: (i) correct responses (+35 points), (ii) incorrect responses (-20 points), and (iii) missed trials (--50 points), accompanying with a sound and color of the feedback text according to the outcome.
After receiving feedback, participants are asked to press the space bar to start the next trial in their own time (Fig.~\ref{S3}). All displays and data collection are coded using the MATLAB software and Psychophysics Toolbox Version 3 (PTB-3)~\cite{kleiner2007s}. 

\textbf{Control experiment.} Among the participants of Experiment A, nine individuals have also participated in the control experiment.
For the control experiment, the lab computer has a BenQ ZOWIE XL2731 monitor, with a spatial resolution of [1920 px $\times$ 1080 px], and a temporal resolution of 144 Hz. The temporal resolution in the motor implementation experiment is set the same as in Experiment A.
The control experiment comprises three parts with varying parameters of the stimuli. For all the parts, participants are instructed to press the right (left) key as soon as the disk reaches the right (left) red line visible to the participants throughout the trials. 
Note that here the participants do not make decisions on the direction of the disk's motion. 
Instead, they only need to react as soon as the disk reaches either of the visible lines. Therefore, we refer to the control experiment as a decision-free response task.
All parts of the control experiment consist of 160 trials separated into 4 blocks. The first part (`no diffusion' version) has stimuli with a velocity of 480 px/s. In the second part of the experiment (`$Y$-axis diffusion' version), in addition to the drift, there is also a diffusivity component in the $y$-axis of 21600 px$^2$/s. The third part (`original parameter' version) has stimuli with the same velocity and diffusivity as Experiment A. The order of these parts are randomized within participants. In the control experiment, participants do not receive any feedback.

\section{Linear mixed effects statistical modeling}
\label{sec:lin_mix_eff}

In this section, we provide additional details about the statistical tests that we carry out to support the data analyses presented in the Main Text. In each subsection, we discuss the statistical tests for the corresponding metric (decision thresholds, accuracies, and times) first for Experiment A and then for Experiment B.

\textbf{Decision thresholds}. \par

The trial-by-trial $X_{\rm dec}$ data are transformed as sqrt(constant-abs($X_{\rm dec}$)) due to the left-skewed distribution before taking the average for each condition and participant.

\textit{Experiment A.} We consider $\langle X_{\rm dec} \rangle$ as a function of experimental conditions (condition) as factor,  and participant identities (id) as random effect to obtain $\rm{model}_A (\textit{X}_{\rm dec})$ (marginal $R^2$:  0.007, conditional $R^2$: 0.83):
\begin{align}
\langle\textit{X}_{\rm dec} \rangle \, & \sim  \text{condition} + (1 \mid \text{id}).
\label{eq:mix-eff-mod-xdec}
\end{align}
ANOVA type-III analysis performed on the obtained  model $X_{\rm dec}$, reveals no significant effect $F_{3,60}$ = 1.17, $p =$ 0.3289, $\eta^2$ = 0.06. The mean decision thresholds are kept fixed under different experimental conditions, meaning that the participants do not modulate their mean threshold with different conditions.

 \textit{Experiment B.} In a subsequent analysis on Experiment B, we fit another linear mixed-effects model to assess the effects of $v$, $D$, and their interaction on $\langle X_{\rm dec} \rangle$, accounting for the random effects due to within participants design. $\text{Model}_{\rm B} (X_{\rm dec})$ (marginal $R^2$:  0.014, conditional $R^2$: 0.93):
\begin{align}
\langle X_{\rm dec} \rangle \, & \sim  \text{$v$*$D$} + (1 \mid \text{id}).
\label{eq:mix-eff-mod-xdec2}
\end{align}

ANOVA type-III analysis performed on the obtained  model given the fixed effects and their interaction, reveals significant effects for the fixed effect of $v$ with $F_{2,184}$ = 14.79, $p <$ 0.001,  $\eta^2$ = 0.14 and for $D$ with $F_{2,184}$ = 5.14, $p <$ 0.01,  $\eta^2$ = 0.05; yet does not reveal a significant effect for their interaction $F_{4,184}$ = 0.77, $p =$ 0.55. 

Multiple comparisons on pairs of $v$ show a significant difference between low $v$ and medium $v$ with $t_{184}$ = 4.38, $p =$ 0.0001, and high $v$ with $t_{184}$ = 4.98, $p <$ 0.0001, suggesting an increase in $X_{\rm dec}$ as $v$ increases from lowest to highest (see Table~\ref{tab:LMER_xdec_vel}). Comparisons between pairs of $D$ reveal a significant difference between the low and medium $D$ with $t_{184}$ = -3.21, $p =$ 0.0048, suggesting a decrease in $X_{\rm dec}$ in the medium group $D$ compared to the low group $D$ (see Table~\ref{tab:LMER_xdec_diff}).

\begin{table}[ht]
\centering
\begin{tabular}{lrrrrl}
  \hline
 & estimate & SE & df & t stat & p value \\ 
  \hline
low $v$ - medium $v$ & 0.326 & 0.074 & 184 & 4.381 & 0.0001 \\ 
low $v$ - high $v$ & 0.371 & 0.074 & 184 & 4.981 & $<$.0001 \\ 
medium $v$  - high $v$  & 0.045 & 0.074 & 184 & 0.600 & 1.0000 \\ 
   \hline
\end{tabular}
\caption{Multiple comparisons of velocity pairs for $\rm{Model}_B (\textit{X}_{\rm dec})$. Degrees-of-freedom method: kenward-roger. P-value adjustment: bonferroni method for 3 tests.}
\label{tab:LMER_xdec_vel}
\end{table}

\begin{table}[ht]
\centering
\begin{tabular}{lrrrrl}
  \hline
 & estimate & SE & df & t stat & p value \\ 
  \hline
low $D$ - medium $D$ & -0.239 & 0.074 & 184 & -3.206 & 0.0048 \\ 
 low $D$ - high $D$ & -0.115 & 0.074 & 184 & -1.548 & 0.3698 \\ 
  medium $D$ - high $D$ & 0.123 & 0.074 & 184 & 1.658 & 0.2972 \\ 
   \hline
\end{tabular}
\caption{ Multiple comparisons of diffusivity pairs for $\rm{Model}_B (\textit{X}_{\rm dec})$. Degrees-of-freedom method: kenward-roger. P-value adjustment: bonferroni method for 3 tests.}
\label{tab:LMER_xdec_diff}
\end{table}

\textbf{Accuracies ($\alpha$). } \par
\textit{Experiment A.} Similar to the decision thresholds analysis, we obtain the model accuracy from the accuracy data by considering experimental conditions as factor and participant identities as random effect, $\rm{model}_A (\alpha)$ (marginal $R^2$:  0.25, conditional $R^2$: 0.74) :
\begin{align}
\alpha &\sim \text{conditions}+ (1 \mid \text{id}).
\label{eq:mix-eff-mod-accu}
\end{align}
ANOVA type-III analysis performed on the obtained model accuracy reveals a statistically significant effect of the experimental conditions over accuracy with $F_{3,60}$ = 25.94, $p <$ 0.001,  $\eta^2$ = 0.56. 
Post hoc analysis reveals that all comparisons are significantly different (all  $t_{60}$ $>$ 2.83,  all $p <$ 0.0377) except for the comparison between condition I and IV (see Table~\ref{tab:LMER_acc_cond}). 

\begin{table}[ht]
\centering
\begin{tabular}{lrrrrl}
  \hline
& estimate & SE & df & t stat& p value\\ 
  \hline
I - II& 0.045 & 0.008 & 60 & 5.804 & $<$.0001 \\ 
  I - III& -0.022 & 0.008 & 60 & -2.832 & 0.0377 \\ 
  I - IV& 0.011 & 0.008 & 60 & 1.432 & 0.9447 \\ 
  II - III& -0.067 & 0.008 & 60 & -8.636 & $<$.0001 \\ 
  II - IV& -0.034 & 0.008 & 60 & -4.372 & 0.0003 \\ 
  III - IV& 0.033 & 0.008 & 60 & 4.264 & 0.0004 \\ 
   \hline
\end{tabular}
\caption{Multiple comparisons of condition pairs for $\rm{Model}_A (\alpha)$. Degrees-of-freedom method: kenward-roger. P-value adjustment: bonferroni method for 6 tests.}
\label{tab:LMER_acc_cond}
\end{table}

 \textit{Experiment B.} Another linear-mixed effect model of accuracy is performed using Experiment B data, in order to further observe the separate effects of $v$ and $D$ over accuracy. $\rm{Model}_B (\alpha)$ (Marginal $R^2$: 0.84 and conditional $R^2$: 0.93):
\begin{align}
\alpha &\sim \text{ $v$*$D$} + (1 \mid \text{id}) .
\label{eq:mix-eff-mod-accu2}
\end{align}
ANOVA type-III analysis performed on the obtained model accuracy reveals statistically significant effects for the fixed effect $v$ with $F_{2,184}$ = 564.85, $ p <$ 0.001, $\eta^2$ = 0.86, $D$ with $F_{2,184}$ = 707.54, $ p <$ 0.001, $\eta^2$ = 0.88 and their interaction over accuracy with $F_{4,184}$ = 40.98, $ p <$ 0.001, $\eta^2$ = 0.47. Furthermore, multiple comparisons indicate that when conditions grouped by velocity, all the diffusivity pairs are significantly different except one (see Table~\ref{tab:LMER_acc_interaction_vel}) with a drop in accuracy for larger values of diffusivity (all  $t_{184}$ $>$ 3.64, all $p <$ 0.001). Similarly all multiple comparisons across velocity pairs when grouped by diffusivity are significant except one (all  $t_{184}$ $>$ 3.19, all $p <$ 0.005), indicating an overall increase in the accuracy with increasing velocity (see Table~\ref{tab:LMER_acc_interaction_diff} for grouping by diffusivity).

\begin{table}[ht]
\centering
\begin{tabular}{lrrrrl}
  \hline
 & estimate & SE & df & t stat & p value \\ 
  \hline
\multicolumn{6}{l}{$v$ = low}\\
low $D$ - medium $D$ & 0.101 & 0.008 & 184 & 13.391 & $<$.0001 \\ 
low $D$ - high $D$ & 0.096 & 0.008 & 184 & 12.692 & $<$.0001 \\ 
medium $D$ - high $D$ & -0.005 & 0.008 & 184 & -0.699 & 1.0000 \\ 
   \hline
\multicolumn{6}{l}{$v$ = medium}\\
low $D$ - medium $D$ & 0.109 & 0.008 & 184 & 14.341 & $<$.0001 \\ 
low $D$ - high $D$ & 0.136 & 0.008 & 184 & 17.985 & $<$.0001 \\ 
medium $D$ - high $D$ & 0.028 & 0.008 & 184 & 3.644 & 0.0010 \\ 
   \hline
\multicolumn{6}{l}{$v$ = high}\\
low $D$ - medium $D$ & 0.174 & 0.008 & 184 & 22.987 & $<$.0001 \\ 
low $D$ - high $D$ & 0.228 & 0.008 & 184 & 30.102 & $<$.0001 \\ 
medium $D$ - high $D$ & 0.054 & 0.008 & 184 & 7.115 & $<$.0001 \\ 
   \hline
\end{tabular}
\caption{ Multiple comparisons of diffusivity pairs grouped by velocity for $\rm{Model}_B (\alpha)$. Degrees-of-freedom method: kenward-roger. P-value adjustment: bonferroni method for 3 tests.}
\label{tab:LMER_acc_interaction_vel}
\end{table}

\begin{table}[ht]
\centering
\begin{tabular}{lrrrrl}
  \hline
 & estimate & SE & df & t stat & p value \\ 
  \hline
\multicolumn{6}{l}{$D$ = low}\\
low $v$ - medium $v$ & -0.120 & 0.008 & 184 & -15.881 & $<$.0001 \\ 
low $v$ - high $v$ & -0.210 & 0.008 & 184 & -27.717 & $<$.0001 \\ 
medium $v$ - high $v$ & -0.090 & 0.008 & 184 & -11.836 & $<$.0001 \\ 
   \hline
\multicolumn{6}{l}{$D$ = medium}\\
low $v$ - medium $v$ & -0.113 & 0.008 & 184 & -14.931 & $<$.0001 \\ 
low $v$ - high $v$ & -0.137 & 0.008 & 184 & -18.121 & $<$.0001 \\ 
medium $v$ - high $v$ & -0.024 & 0.008 & 184 & -3.190 & 0.0050 \\ 
   \hline
\multicolumn{6}{l}{$D$ = high}\\
low $v$ - medium $v$ & -0.080 & 0.008 & 184 & -10.588 & $<$.0001 \\ 
low $v$ - high $v$ & -0.078 & 0.008 & 184 & -10.307 & $<$.0001 \\ 
medium $v$ - high $v$ & 0.002 & 0.008 & 184 & 0.281 & 1.0000 \\ 
   \hline
\end{tabular}
\caption{ Multiple comparisons of velocity pairs grouped by diffusivity for $\rm{Model}_B (\alpha)$. Degrees-of-freedom method: kenward-roger. P-value adjustment: bonferroni method for 3 tests.}
\label{tab:LMER_acc_interaction_diff}
\end{table}

\textbf{Decision times.} \par
\textit{Experiment A.} We consider transformed $T_{\rm dec}$ data (-constant/$T_{\rm dec}$) as a function of experimental conditions as factor, and participant identities as random effect to obtain $\rm{model}_A (\textit{T}_{\rm dec})$  (marginal $R^2$:  0.52, conditional $R^2$: 0.92): 
\begin{align}
\langle\textit{T}_{\rm dec} \rangle \
&\sim \text{conditions} + 
(1 \mid \text{id} ) .
\label{eq:mix-eff-mod-tdec}
\end{align}
Furthermore, with the obtained model $\langle T_{\rm dec} \rangle$ we perform the ANOVA type-III analysis, which reveals a statistically significant effect of the experimental conditions over $T_{\rm dec}$ with $F_{3,60}$ = 170.36, $ p <$ 0.001, $\eta^2$ = 0.89. Post hoc analysis reveals that all comparisons are significantly different (all  $t_{60}$ $>$ 3.89,  all $p <$ 0.0015), as $\Sigma$ increases from condition I to IV,  $T_{\rm dec}$ gets faster (see Table~\ref{tab:LMER_tdec_cond}). 

\begin{table}[ht]
\centering
\begin{tabular}{lrrrrl}
  \hline
& estimate & SE & df & t stat& p value\\ 
  \hline
I - II& 0.073 & 0.019 & 60 & 3.889 & 0.0015 \\ 
  I - III& 0.178 & 0.019 & 60 & 9.431 & $<$.0001 \\ 
  I - IV& 0.401 & 0.019 & 60 & 21.198 & $<$.0001 \\ 
  II - III& 0.105 & 0.019 & 60 & 5.541 & $<$.0001 \\ 
  II - IV& 0.327 & 0.019 & 60 & 17.308 & $<$.0001 \\ 
  III - IV& 0.222 & 0.019 & 60 & 11.767 & $<$.0001 \\ 
   \hline
\end{tabular}
\caption{Multiple comparisons of condition pairs for $\rm{Model}_A (\textit{T}_{\rm dec})$. Degrees-of-freedom method: kenward-roger. P-value adjustment: bonferroni method for 6 tests.}
\label{tab:LMER_tdec_cond}
\end{table}

 \textit{Experiment B.} We also perform a model of the log-transformed $T_{\rm dec}$ data of Experiment B. $\rm{Model}_B (\textit{T}_{\rm dec})$ (Marginal $R^2$: 0.59, conditional $R^2$: 0.96):
\begin{align}
\langle \textit{T}_{\rm dec}\rangle  &\sim \text{$v$*$D$} + (1 \mid \text{id}).
\label{eq:mix-eff-mod-tdec2}
\end{align}
ANOVA type-III analysis performed on the obtained model $\langle T_{\rm dec} \rangle$ reveals statistically significant effects for the fixed effect $v$ ($F_{2,184}$ = 22.10, $ p <$ 0.001, $\eta^2$ = 0.19), the fixed effect $D$ ($F_{2,184}$ = 1551.72, $ p <$ 0.001, $\eta^2$ = 0.94); and their interaction over $\langle T_{\rm dec} \rangle$ ($F_{4,184}$ = 33.89, $ p <$ 0.001, $\eta^2$ = 0.42). This indicates a stronger effect of $D$ compared to $v$ on the dependent variable. Furthermore, post-hoc comparisons indicate that when conditions grouped by velocity, all the diffusivity pairs are significantly different (all  $t_{184}$ $>$ 7.87,  all $p <$ 0.0001) (see Table~\ref{tab:LMER_tdec_interaction_vel}) with a decrease of $T_{\rm dec}$ for larger values of diffusivity (see also Table~\ref{tab:LMER_tdec_interaction_diff} for grouping by diffusivity).  

\begin{table}[ht]
\centering
\begin{tabular}{lrrrrl}
  \hline
 & estimate & SE & df & t stat & p value \\ 
  \hline
\multicolumn{6}{l}{$v$ = low}\\
low $D$ - medium $D$ & 0.43 & 0.02 & 184 & 21.758 & $<$.0001 \\ 
low $D$ - high $D$ & 0.73 & 0.02 & 184 & 37.258 & $<$.0001 \\ 
medium $D$ - high $D$ & 0.31 & 0.02 & 184 & 15.500 & $<$.0001 \\ 
   \hline
\multicolumn{6}{l}{$v$ = medium}\\
low $D$ - medium $D$ & 0.25 & 0.02 & 184 & 12.648 & $<$.0001 \\ 
low $D$ - high $D$ & 0.66 & 0.02 & 184 & 33.723 & $<$.0001 \\ 
medium $D$ - high $D$  & 0.42 & 0.02 & 184 & 21.075 & $<$.0001 \\ 
   \hline
\multicolumn{6}{l}{$v$ = high}\\
low $D$ - medium $D$ & 0.35 & 0.02 & 184 & 17.538 & $<$.0001 \\ 
low $D$ - high $D$ & 0.50 & 0.02 & 184 & 25.412 & $<$.0001 \\ 
medium $D$ - high $D$  & 0.16 & 0.02 & 184 & 7.874 & $<$.0001 \\ 
   \hline
\end{tabular}
\caption{ Multiple comparisons of diffusivity pairs grouped by velocity for $\rm{Model}_B (\textit{T}_{\rm dec})$. Degrees-of-freedom method: kenward-roger. P-value adjustment: bonferroni method for 3 tests.}
\label{tab:LMER_tdec_interaction_vel}
\end{table}

\begin{table}[ht]
\centering
\begin{tabular}{lrrrrl}
  \hline
 & estimate & SE & df & t stat & p value \\ 
  \hline
\multicolumn{6}{l}{$D$ = low}\\
low $v$ - medium $v$ & 0.11 & 0.02 & 184 & 5.676 & $<$.0001 \\ 
low $v$ - high $v$ & 0.18 & 0.02 & 184 & 9.160 & $<$.0001 \\ 
medium $v$ - high $v$ & 0.07 & 0.02 & 184 & 3.484 & 0.0019 \\ 
   \hline
\multicolumn{6}{l}{$D$ = medium}\\
low $v$ - medium $v$ & -0.07 & 0.02 & 184 & -3.434 & 0.0022 \\ 
low $v$ - high $v$ & 0.10 & 0.02 & 184 & 4.940 & $<$.0001 \\ 
medium $v$ - high $v$ & 0.17 & 0.02 & 184 & 8.374 & $<$.0001 \\ 
   \hline
\multicolumn{6}{l}{$D$ = high}\\
low $v$ - medium $v$ & 0.04 & 0.02 & 184 & 2.141 & 0.1008 \\ 
low $v$ - high $v$ & -0.05 & 0.02 & 184 & -2.686 & 0.0237 \\ 
medium $v$ - high $v$ & -0.10 & 0.02 & 184 & -4.827 & $<$.0001 \\ 
   \hline
\end{tabular}
\caption{ Multiple comparisons of velocity pairs grouped by diffusivity for $\rm{Model}_B (\textit{T}_{\rm dec})$. Degrees-of-freedom method: kenward-roger. P-value adjustment: bonferroni method for 3 tests.}
\label{tab:LMER_tdec_interaction_diff}
\end{table}

\section{Mean decision time in  Wald's SPRT}
\label{app:theo}

We provide here the analytical calculation of the mean decision time on the direction of time's arrow for the dynamics of a 1D drift diffusion process
\begin{equation}
    \dot{X}_t = v + \sqrt{2D}\xi_t,
    \label{eq:a1}
\end{equation}
with $\xi_t$ a Gaussian white noise with zero mean $\langle\xi_t \rangle =0$ and autocorrelation $\langle\xi_{t}\xi_{t'}\rangle = \delta (t-t')$, see also Ref.~\cite{roldan2015decision}.  The task by such Wald SPRT is to discern as soon as possible, with a prescribed accuracy whether a stochastic trajectory $X_{[0,t]}$ (sequence of data) is produced by the dynamics given by Eq.~\eqref{eq:a1} (hypothesis $H_1$) or by the time-reversed dynamics (hypothesis $H_2$) which in this case reads
\begin{equation}
    \dot{X}_t = - v +\sqrt{2D}\xi_t.
\end{equation}
In the following, we will also assume that the system has periodic boundary conditions and the initial position is set to $X_0=0$.  Under these assumptions, Wald's log likelihood ratio associated with this decision-making process reads
\begin{equation}
    \mathcal{L}_t = \ln \frac{\mathcal{P}(X_{[0,t]}\vert H_1)}{\mathcal{P}(X_{[0,t]}\vert H_2)} = \frac{v}{D} X_t,
    \label{eq:a3}
\end{equation}
see Refs.~\cite{roldan2015decision,roldan2022martingales} for a detailed proof of the second equality in~\eqref{eq:a3}. In what follows, we will also assume that the accuracies $P(D_{\rm dec}=\pm 1\vert H=\pm 1)=\alpha$ are symmetric and both equal to $\alpha > 1/2$, that is, the decision maker performs better than a random guess. 

Under the aforementioned assumptions, this specific Wald's SPRT maps into a first-passage time problem for $\mathcal{L}_t$ to reach any of two absorbing boundaries located at $\lambda$ and $-\lambda$, with $\lambda=\ln [\alpha/(1-\alpha)]$ the dimensionless threshold amplitude, which for SPRT is constant and fixed across trials once specified the accuracy $\alpha$. 
From Eq.~\eqref{eq:a3}, we find that the decision time
\begin{eqnarray}
    T &=& \inf\{t\geq 0 \;\vert\; \mathcal{L}_t\notin (-\lambda, \lambda) \;\}, \\
&=&  \inf\{t\geq 0 \;\vert\; X_t\notin (-\ell, \ell) \;\},   \label{eq:a4}
\end{eqnarray}
with
\begin{equation}
    \ell = \frac{D}{v}\lambda = \frac{D}{v}\ln\left(\frac{\alpha}{1-\alpha}\right),
    \label{eq:ell}
\end{equation}
the threshold amplitude.
In other words, the decision time in Wald's SPRT for this problem is given by the first escape time of a drift-diffusion process from a symmetric interval $(-\ell,\ell)$. 
The mean escape time associated with the first-passage-time problem~\eqref{eq:a4} has been solved analytically in many instances; see, e.g. the reference book~\cite{redner2001guide}. We give here a quick derivation for the sake of completeness, using martingale theory (see Ref.~\cite{roldan2022martingales} for a review).

The process 
\begin{equation}
    M_t = \exp(-vX_t/D),
\end{equation}
with initial value $M_0=1$ is a martingale, i.e. its conditional average conditioned on the history up to time $s<t$ equals to its value at the last time of conditioning, i.e. $\langle M_t \vert X_{[0,s]}\rangle = M_s$. The fact of this process being a martingale implies, among other properties, that its average value at the decision time $T$ given by Eq.~\eqref{eq:a4} obeys Doob's optional stopping theorem~\cite{doob}
\begin{equation}
    \langle M_T \rangle =M_0=1.
    \label{eq:a6}
\end{equation}
Eq.~\eqref{eq:a6} can be rewritten by unfolding the average at the decision time (left hand side) 
\begin{equation}
    \langle M_T\rangle  = P_+\langle M_T \vert + \rangle +  P_-\langle M_T \vert - \rangle ,
    \label{eq:a7}
\end{equation}
where $P_+=\text{Prob}(\mathcal{L}_T = D\ell /v)=\alpha $ and $P_- =\text{Prob}(\mathcal{L}_T =- D\ell /v)=1-\alpha $  are the absorption probability in the positive and negative thresholds, which correspond respectively to the probability to make a right and a wrong decision. From Eqs.~(\ref{eq:a6}-\ref{eq:a7}) and $P_+ + P_- =1$, we get
\begin{eqnarray}
    P_+ &=& \frac{\langle M_T\vert -\rangle -1}{\langle M_T\vert -\rangle -\langle M_T\vert +\rangle }, \label{eq:a8}\\ 
    P_- &=& \frac{1-\langle M_T\vert +\rangle}{\langle M_T\vert -\rangle -\langle M_T\vert +\rangle}.\label{eq:a9}
\end{eqnarray}
Eqs.~(\ref{eq:a8}-\ref{eq:a9}) can be specialized to the process~\eqref{eq:a6} by noting that
\begin{equation}
    \langle M_T\vert +\rangle = \exp(-v\ell/ D),\qquad \langle M_T\vert -\rangle = \exp(v\ell/ D).\label{eq:a11}
\end{equation}
Substituting Eqs.~\eqref{eq:a11} into Eqs.~\eqref{eq:a8} and~\eqref{eq:a9} and after some algebra, we get
\begin{eqnarray}
    P_+ &=& \frac{\exp(v\ell/D)}{1+\exp(v\ell/D)  },\label{eq:a13}\\
    P_- &=& \frac{1}{1+\exp(v\ell/D) }.\label{eq:a14}
\end{eqnarray}
Substituting in the above equations the value of the threshold amplitude $\ell$ in Eq.~\eqref{eq:ell} one retrieves $P_+=\alpha$ and $P_-=1-\alpha$, as expected. In other words, the accuracy reads
\begin{equation}
    \alpha = \frac{1}{1+\exp(-v\ell/D) }.\label{eq:alpha}
\end{equation}
We also note here that the accuracy in this Wald SPRT depends only in the parameter values through the P\'eclet number associated with the drift diffusion process in the interval $(-\ell,\ell)$ which for this example is given by
\begin{equation}
    \text{Pe}=\frac{v\ell}{D}.
\end{equation}

The analytical expressions for the absorption probabilities~(\ref{eq:a13}-\ref{eq:a14}) allow us to compute the mean decision time in Wald's SPRT. First we note that we can write the average value of the particle position at the decision time explicitly as
\begin{equation}
    \langle X_T\rangle  = P_+ \ell  - P_- \ell .\label{eq:a16}
\end{equation}
This result follows from noting that the decision-making process ends in $\ell$ with probability $P_+$ and in $-\ell$ with probability $P_-$. On the other hand, we also have
\begin{eqnarray}
    \langle X_T\rangle =v \langle T\rangle  + \sqrt{2D}\langle B_T\rangle = v\langle T\rangle ,\label{eq:a17}
\end{eqnarray}
where $B_T$ denotes here the Wiener process at time $T$. In Eq.~\eqref{eq:a17}, the first equality follows from the equation of motion of the process~\eqref{eq:a1} and the second equality from Doob's optional stopping theorem~\eqref{eq:a6} applied to the Wiener process, which is a martingale. Combining Eqs.~\eqref{eq:a13},~\eqref{eq:a14},~\eqref{eq:a16}, and~\eqref{eq:a17} we get
\begin{equation}
    \langle T\rangle = \frac{\ell}{v} \tanh\left( \frac{v\ell}{2D}\right).
    \label{eq:a19}
\end{equation}
Eq.~\eqref{eq:a19} together with the steady-state rate of entropy production~\cite{roldan2015decision}
\begin{equation}
    \Sigma = \langle \dot{\mathcal{L}}_t\rangle  =\frac{v\langle \dot{X}_t\rangle}{D}  = \frac{v^2}{D}\label{eq:d21} ,
\end{equation}
yields
\begin{equation}
    \Sigma \expval{T}=
    \mathrm{Pe} \, \tanh \qty(\frac{\mathrm{Pe}}{2}) ,
    \label{eq:MDTwald}
\end{equation}
which holds for Wald's SPRT. For suboptimal decision making protocols, it is shown in Ref.~\cite{roldan2015decision} that the mean decision time $\langle T_{\rm dec}\rangle$ exceeds Wald's SPRT mean decision time~\eqref{eq:MDTwald}, which implies Eq.~\eqref{eq:TURdt} in the Main Text.
 
\section{Chi-square analysis for Wald's SPRT}
\label{chi_sup}

\textbf{Accuracies.} Chi-square (${\chi}^{2}$) goodness-of-fit tests are performed for each condition to determine to what extent the distribution of accuracies predicted from SPRT is close to the distribution of the experimental accuracy data. The ${\chi}^{2}_c$ of accuracy for condition $c$ is computed using
\begin{align}
{\chi}^{2}_{c} & = 
 \frac{\qty({{\langle {\alpha}_{\mathrm{SPRT}}^{p,c} \rangle _{p}} -
 {\langle \alpha^{p,c} \rangle _{p}} })^{2} }{\mathrm{std}({\alpha}_{\mathrm{SPRT}}^{p,c})^{2}_{p} +  {\mathrm{std}}({\alpha}^{p,c})^{2}_{p} }
\label{eq:chiAcc},
\end{align}
where $\alpha^{p,c}$ and ${\alpha}_{\mathrm{SPRT}}^{p,c}$ denote the accuracies from the experiment and the predictions of SPRT for participant $p$ in condition $c$, respectively. Here, the symbols $\expval{\cdot}_p$ and $\mathrm{std}(\cdot)_p$ represent the average and standard deviation with respect to the participants, respectively.
The corresponding ${\chi_c}$ values for conditions I, II, III and IV are obtained to be 0.68, 0.42, 1.62 and 0.43, respectively. These values indicate the standard deviation of the difference between theoretical and experimental data of accuracy in comparison to a distribution with zero mean.

\textbf{Decision times.} Chi-square (${\chi}^{2}$) goodness-of-fit tests are also performed for each condition and each participant to determine to what extent the distribution of decision times predicted from SPRT is close to the distribution of the experimental decision times data. The ${\chi}^{2}_{p,c}$ of decision times for participant $p$ in condition $c$ is computed using
\begin{align}
{\chi}^{2}_{p,c} & = 
 \qty(\frac{{{\langle {T_{\mathrm{dec}}}\rangle_{\mathrm{SPRT}}^{p,c} } -
 {\langle T_{\mathrm{dec}}^{p,c,i} \rangle _{i}} } }{\mathrm{std}({T_{\mathrm{dec}}}^{p,c,i})_{i}})^{2}
\label{eq:chitimes} ,
\end{align}
where $T_{\mathrm{dec}}^{p,c,i}$ denotes the experimental decision times for participant $p$ in condition $c$ and trial $i$, and $\langle {T_{\mathrm{dec}}}\rangle_{\mathrm{SPRT}}^{p,c}$ denotes the SPRT predictions of mean decision times for participant $p$ in condition $c$. Here, the symbols $\expval{\cdot}_i$ and $\mathrm{std}(\cdot)_i$ represent the average and standard deviation with respect to the trials.
The obtained ${\chi}^{2}_{p,c}$ values for each condition are summed across the participants and then a ${\chi}_c = \sqrt{\sum_{p} \chi^2_{p,c}}$ is calculated. Thus, we obtain $\chi_c$ values 1.16, 0.62, 0.83 and 1.19 for conditions I, II, III, and IV, respectively. These values indicate the standard deviation of the difference between the distributions of experimental decision times and the SPRT's predictions.

\section{Evidence Integration Model}
\label{EIM_sup}

\textbf{Discrete-time interpretation}. One may rewrite Eq.~\eqref{eq:Wtdef_backward} in the Main Text in discrete times with time steps $\Delta t$ as
\begin{align}
{W}_{t + \Delta t} & = {\mathrm{e}}^{-\Delta t/\tau_m} W_{t}  + \frac{\Delta t}{\tau_{m}} X_{t + \Delta t}
 \label{eq:Wtdiscr} .
\end{align}
The first term on the right-hand side of Eq.~\eqref{eq:Wtdiscr} denotes the prior knowledge $W_t$ accumulated until time $t$ and captures the memory of the evidence accrued so far. The exponential factor ${\mathrm{e}}^{-\Delta t/\tau_m}$ in this term acts as a weight, suggesting that the importance of memory increases as the relaxation time $\tau_m$ increases.
On the other hand, the second term on the right-hand side of Eq.~\eqref{eq:Wtdiscr} acts as instantaneous evidence and adds the latest information to the existing memory in terms of the present position of the disk.  
The factor $(\Delta t/ \tau_m)$ in this term suggests that the latest instantaneous evidence for a long relaxation time $\tau_m$ becomes less significant compared to the evidence accumulated previously. 
Thus, by construction, the decision accumulator $W_t$, depending on $\tau_m$, automatically captures a trade-off between memory and instantaneous evidence uptake.
Higher reliance on memory for stimuli further from equilibrium may be a strategy used by participants in order to capture evidence history more holistically and thus reduce the noise of the input in memory.

\begin{figure}[!htbp]
\centering
\includegraphics[scale=0.38]{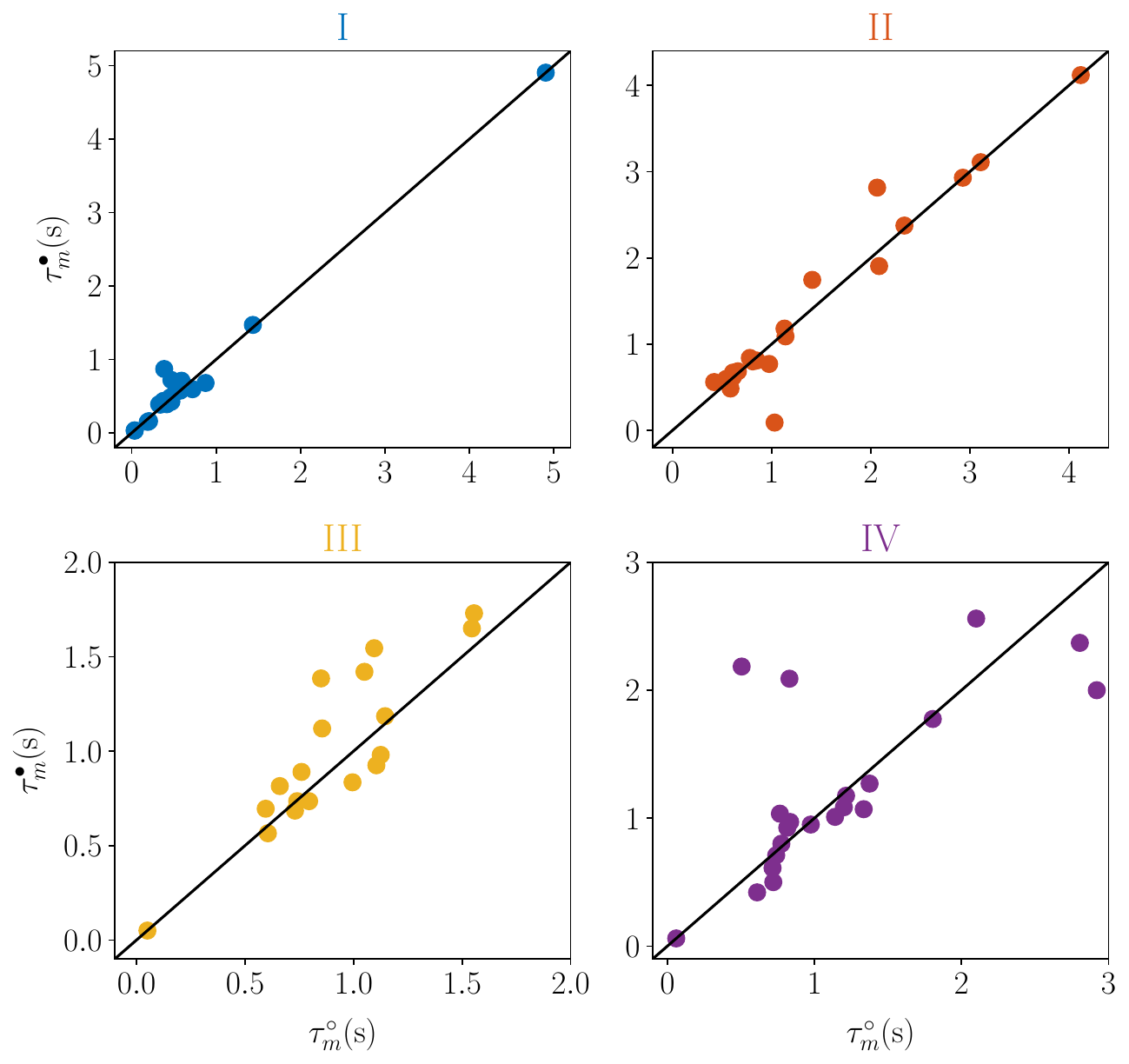}
\caption{\textbf{Cross validation of the memory relaxation time.}  Conditions I (top left), II (top right), III (bottom left), and IV (Experiment A). We divide the trial-by-trial data into two equal-sized parts at random and find two memory relaxation times $\tau_m^*$ and $\tau_m^\diamond$  that accurately matches the average decision time for each respective portion of the experimental data. Points correspond to a scatter plot of these two parameters for each participant and lines with slope one to the ideal best fit.}\label{figS10}
\end{figure}

\textbf{Intrinsic memory relaxation time.} Here, we check the robustness of our assumption that each participant builds for each condition an intrinsic $\tau_m^{\star}$. To do this, we randomly split the trial-by-trial data for each participant into two segments of equal size.
We then identify for each segment the corresponding memory relaxation times denoted by $\tau_m^{\bullet}$ and $\tau_m^\circ$ using the procedure mentioned in the Main Text. 
Fig.~\ref{figS10} displays for each condition the scatter plots of $\tau_m^{\bullet}$ and $\tau_m^\circ$, in which each data point represents a participant.
From the figure, we observe that most of the points lie on the diagonal, indicating the fact that $\tau_m^{\bullet} \simeq \tau_m^{\circ}$.
The similarity of the values of the two independent segments suggests that there exists an intrinsic memory relaxation time associated with each participant under each condition.

\begin{figure}[!htbp]
\centering
\includegraphics[scale=0.35]{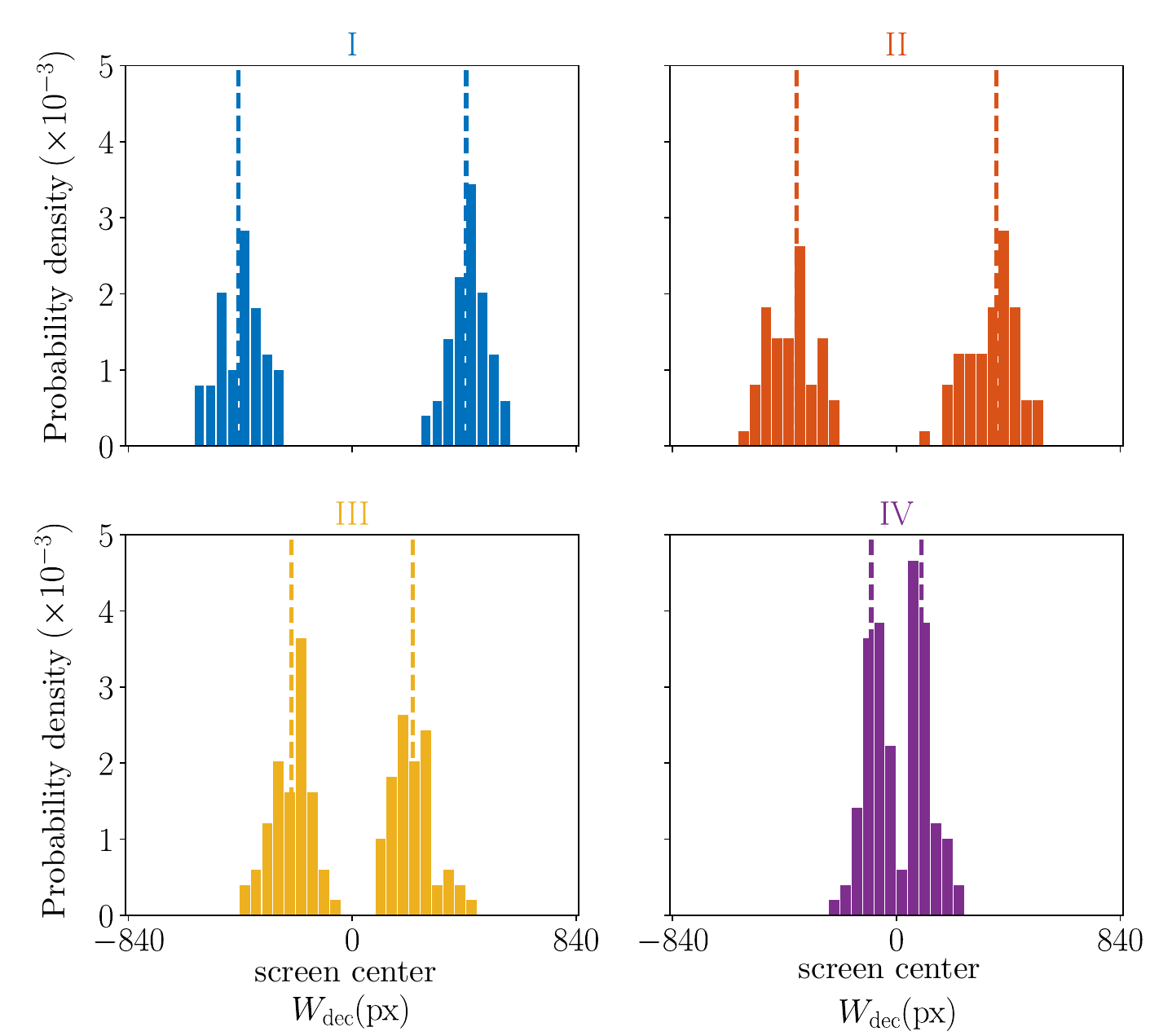}
\caption{\textbf{Probability density of $W_{\rm dec}=W_{T_{\rm dec}}$ in Experiment A.} The histograms show the probability density of $W_{\rm dec}$ obtained from the trial-by-trial analysis of a single participant using Eq.~\eqref{eq:Wtdiscr} across experimental conditions. The dashed lines in each condition display the corresponding  mean values $-L^{\star}_W$ and $L^{\star}_W$.}
\label{figS8}
\end{figure}

\textbf{Intrinsic decision threshold.} We also compute the corresponding $W_{\rm dec}$ that generates $L_W^\star$ on a trial-by-trial basis for each participant using Eq.~\eqref{eq:Wtdiscr}. In Fig.~\ref{figS8}, we plot the probability density of $W_{\rm dec}$ for a typical participant in the different conditions. The dashed lines represent the mean decision thresholds $- L_{W}^\star$ and $L_{W}^\star$ with $L_W^{\star} =\expval{|W_{\rm dec}|}$.
Note that the mean decision thresholds of EIM ($L_W^{\star}$) change substantially with experimental conditions.

\begin{figure}[!htbp]
\centering
\includegraphics[scale=0.45]{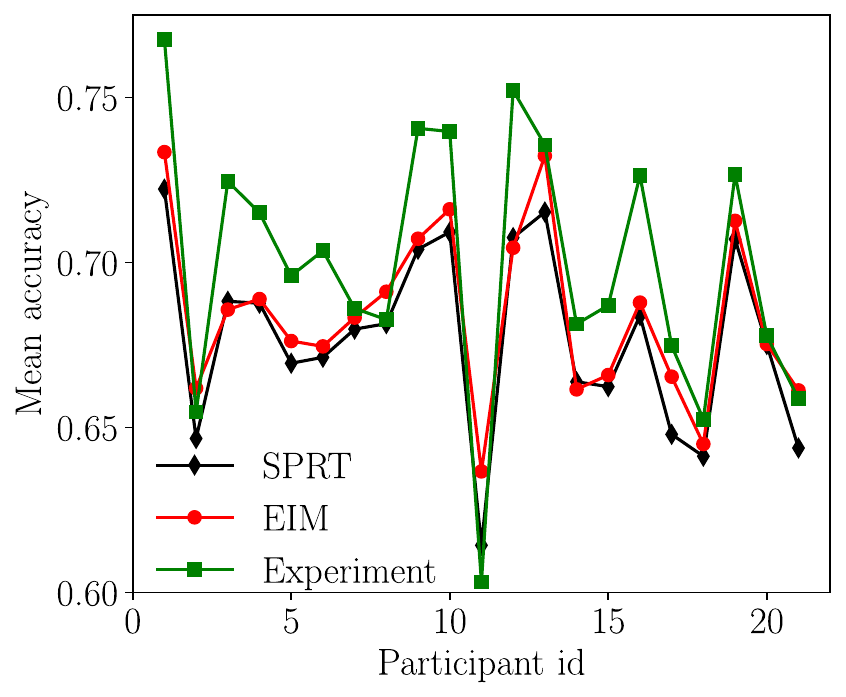}
\caption{\textbf{Participant-by-participant accuracy predictions in Experiment A.} Mean accuracy averaged over all experimental conditions, as a function of the participant's id. The EIM and the SPRT predict similar accuracy, and display correlation across participants.
}
\label{figS4}
\end{figure}

\textbf{Accuracy}. For each participant with its associated $\tau_m^\star$ and $L_{W}^\star$, the EIM overall produces predictions of accuracy similar to those of the SPRT (see  Fig.~\ref{figS4}). 

\begin{figure*}[!htbp]
\centering
\includegraphics[scale=0.3]{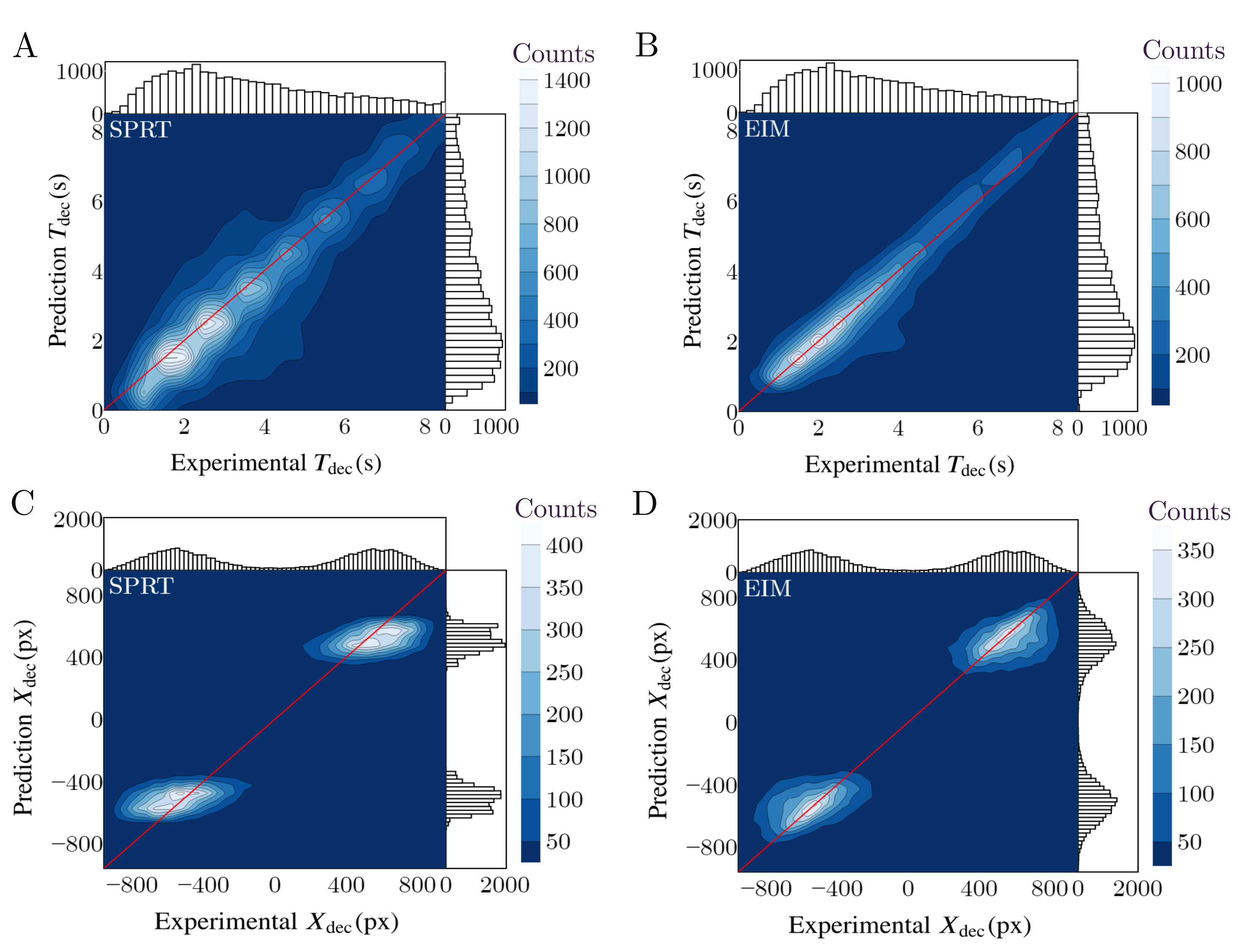}
\caption{ {\bf Comparison of decision-making models (SPRT and EIM) with experimental 
decision times and decision thresholds for Experiment B.} 
(A-B) Joint histogram of the trial-by-trial experimental decision time $T_{\rm dec}$ for all participants and conditions, and the numerical predictions of $T_{\rm dec}$ by Wald's SPRT (A) and the EIM (B). (C-D) Joint histogram of the trial-by-trial experimental decision threshold $X_{\rm dec}$ for all participants and conditions, and the numerical predictions of $X_{\rm dec}$ by Wald's SPRT (C) and the EIM (D). The red lines have slope one, representing the region with perfect agreement between the experimental output and model predictions.}

\label{color_map_exp2}
\end{figure*}

 \textbf{Trial-by-trial predictions in Experiment B.} The Figs.~\ref{color_map_exp2}A-B show joint histograms of the predicted and the experimental decision times collected from all trials of all participants in Experiment B, under all conditions, for the SPRT (Fig.~\ref{color_map_exp2}A) and EIM (Fig.~\ref{color_map_exp2}B). Similarly, Figs.~\ref{color_map_exp2}C-D corresponds to the experimental and predicted decision thresholds at the trial-by-trial level.

\section{Ornstein–Uhlenbeck model}
\label{sec:OU_model}
\begin{figure*}[!htbp]
\centering
\includegraphics[scale=0.5]{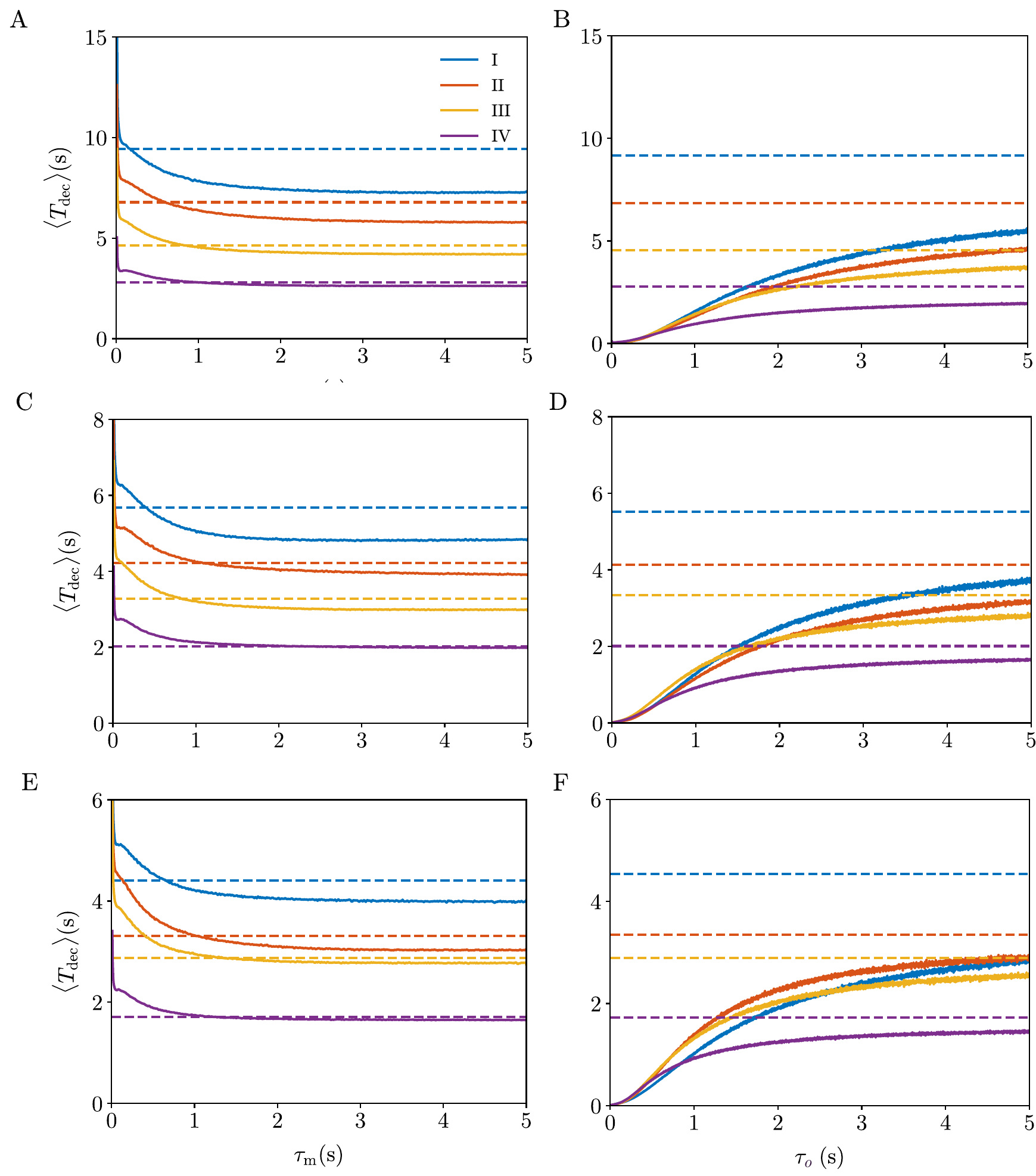}
\caption{ {\bf Inference of the optimal memory relaxation time $\tau_m^\star$ for  EIM (A, C, E) and $\tau_o^\star$ for OU model (B,D,F) in Experiment A}. Three example participants are shown in the four conditions. Each dashed line corresponds to the experimental mean decision time. The solid lines denote the mean escape time of the evidence accumulator $W_t$ from the interval $(-L_W,L_W)$ extracted from the numerical simulations as a function of the corresponding memory relaxation time.}

\label{OU-EIM_comparison}
\end{figure*}

If we replace $X_{t-s}$ with $\dot{X}_{t-s}$ in defining the evidence accumulator $W_t$ (see Eq.~\eqref{eq:Wtdef_backward} in the Main Text), then $W_t$ becomes
\begin{align}
   W_t \equiv \frac{1}{\tau_o} \int_{0}^{t} {\mathrm{d}s} \, \mathrm{e}^{-s/\tau_o} \, \dot{X}_{t-s}\label{eq:Wtdef_OU}
\end{align}
from which one may write $\tau_{o} \dot{W}_{t} = - W_t + \dot{X}_t$. For  $\dot{X}_t=Hv+\sqrt{2D}\xi_t$, the accumulator $W_t$ obeys an Ornstein–Uhlenbeck process
\begin{equation}
       \tau_o \dot{W}_t = -W_t +Hv+\sqrt{2D}\xi_t.
       \label{eq:OU_pro}
\end{equation}
We do the analysis to infer the optimal memory relaxation time $\tau_o^\star$ for each individual for the case of the Ornstein–Uhlenbeck (OU) model,  accumulating evidence through $W_t$ in Eq.~\eqref{eq:OU_pro}. However, we find that there does not exist a set of $\tau_o^\star$ that can fit our experimental data. We show the results of the fitting processes in Fig.~\ref{OU-EIM_comparison}, comparing the EIM (panels on the left) to the OU model (panels on the right) for three example participants.  
We follow an analogous procedure to that used for EIM (see Sec.~\ref{sec:EIM} and Appendix~\ref{EIM_sup}) to the OU model to estimate $\tau_o^\star$ of each participant under each experimental condition. However, it may be seen from the panels on the right that, for these participants, the lines simulated from the OU model for no experimental condition meet the experimental decision times. This observation holds for all participants under all experimental conditions. Thus, the OU model in our experiment does not yield fruitful results. The possible reason for this result is that the velocity in this setting is much less informative than the position of the stimulus. In fact, the OU model is typically used when evidence is an unobservable quantity.

\section{Variance of decision threshold}
\label{sec:var_in_decision_threshold}

\begin{figure*}[!htbp]
\centering
\includegraphics[scale=0.5]{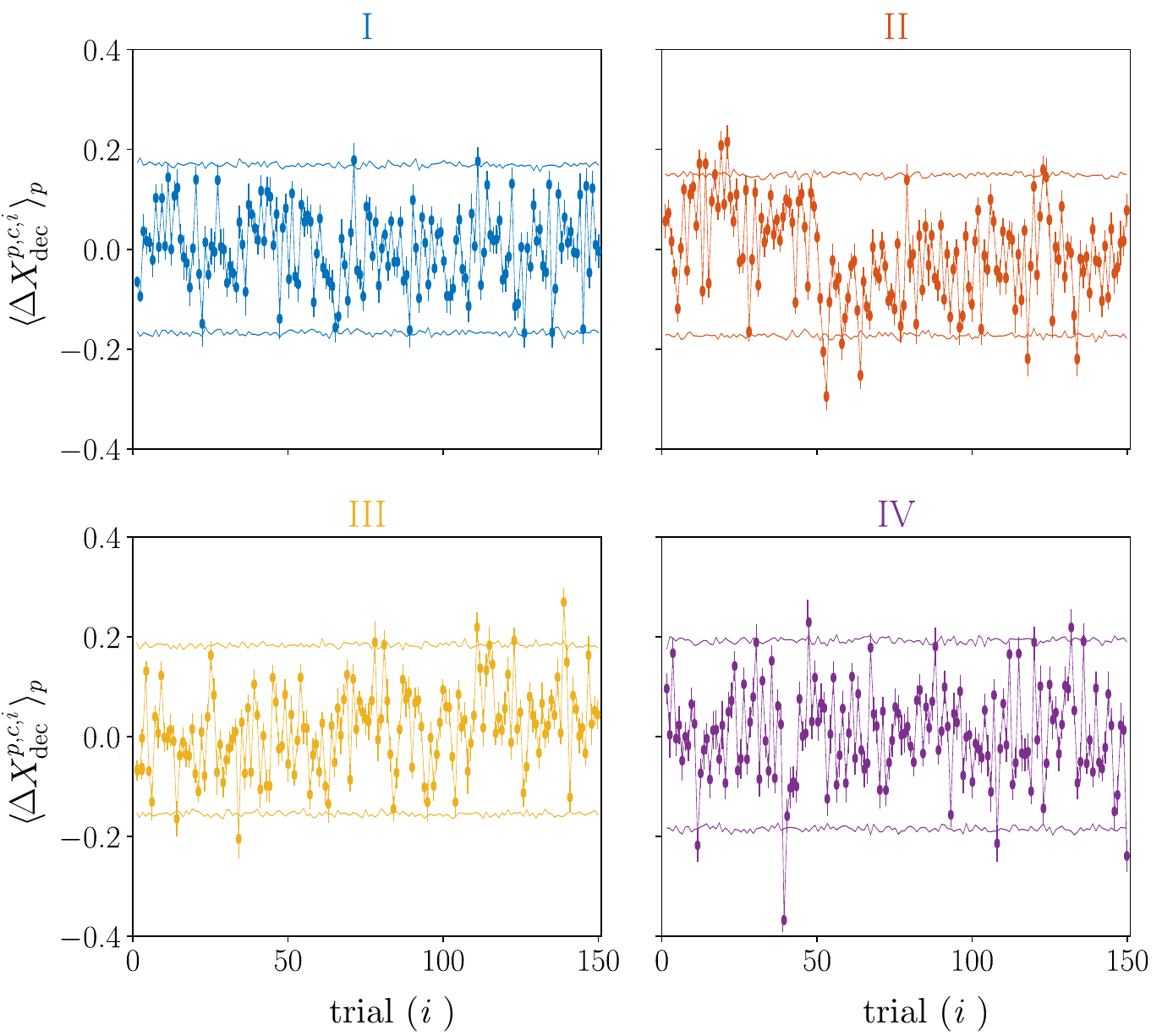}
\caption{\textbf{Participants respond consistently for the same trials.} Given the same trial ($i$),  we calculate the normalized change in each participant's decision threshold with respect to their mean. Each data point represents the mean of the normalized change over all participants in Experiment A. Error bars show SEM. The fluctuating horizontal lines are the $95\%$ confidence intervals.}
\label{fig:S2}
\end{figure*}

We now explore the variance of the decision threshold observed in Figs.~\ref{fig5}C and \ref{fig5}D in the Main Text.
To investigate possible sources of variance, we analyze consistency among participants in Experiment A. 
To proceed, let us denote the observed decision threshold of participant $p$ in trial $i$ for condition $c$ by $X^{p,c,i}_{\mathrm{dec}}$.  For the participant $p$ and condition $c$, we compute in each trial $i$ the relative difference between the decision threshold $X^{p,c,i}_{\mathrm{dec}}$ and its average $\expval{|X_{\mathrm{dec}}^{p,c,j}|}_{j}$ over all trials $j$:
\begin{align}
    \Delta X_{\mathrm{dec}}^{p,c,i} \equiv \frac{\expval{|X_{\mathrm{dec}}^{p,c,j}|}_{j} - |X_{\mathrm{dec}}^{p,c,i}|}{ \expval{|X_{\mathrm{dec}}^{p,c,j}|}_{j} }  .
\end{align}
We take an average of $\Delta X_{\mathrm{dec}}^{p,c,i}$ over the participants, i.e., $\expval{\Delta X_{\mathrm{dec}}^{p,c,i} }_p $, and plot it in Fig.~\ref{fig:S2} as a function of trials ($i$).
We then calculate the null distribution of $\Delta X_{\mathrm{dec}}^{p,c,i}$ by shuffling trials in order to find the 95\% confidence intervals, which are shown as the two fluctuating horizontal lines in the figure panels. The figure shows that the majority of the trials is within the confidence interval. This implies consistency in the decision thresholds between participants. The consistent responses of the participants for the same trial suggest that the variance in decision thresholds comes mainly from the fluctuations of the stimuli.

\begin{figure}[!htbp]
\centering
\includegraphics[scale=0.4]{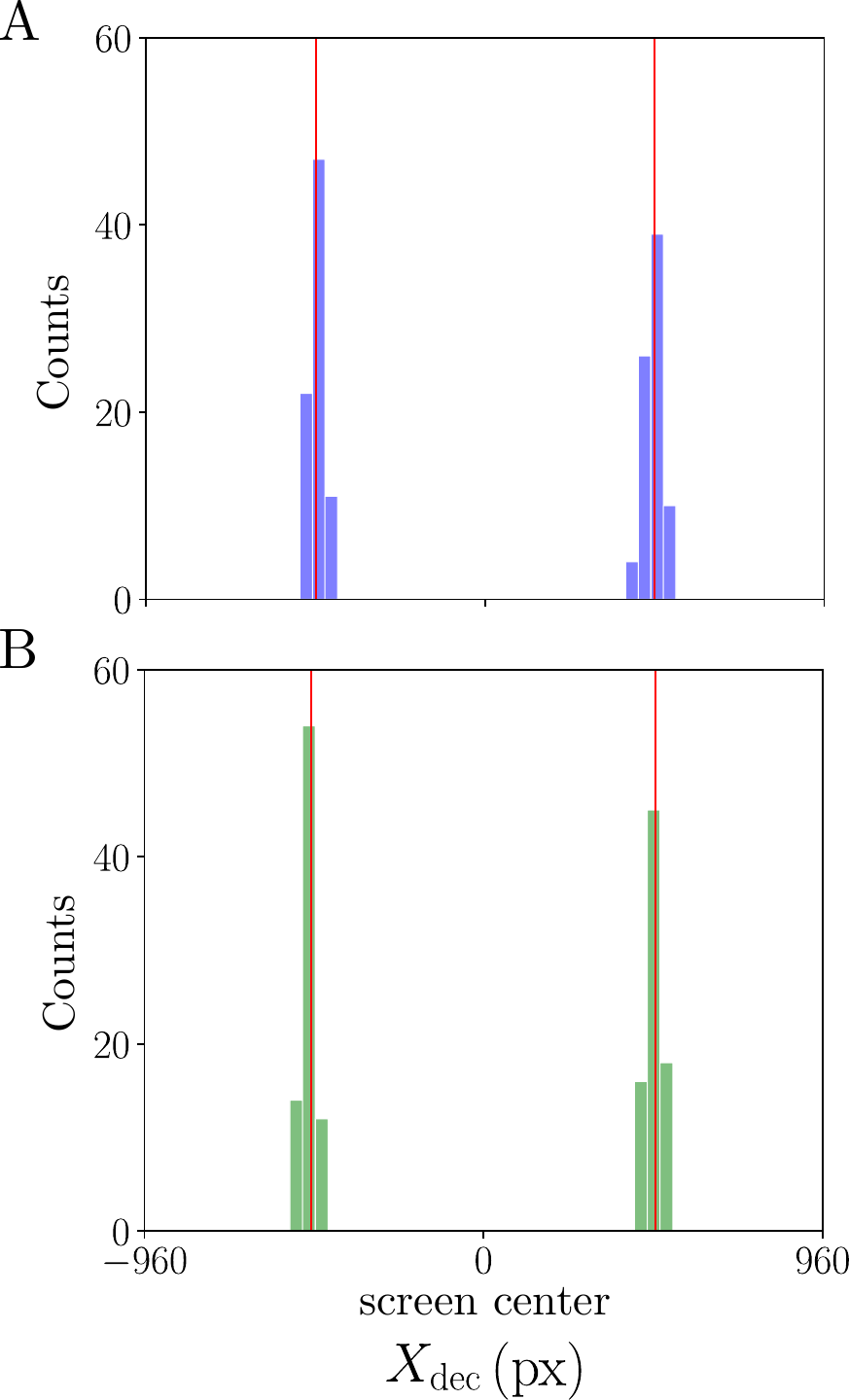}
\caption{\textbf{Control experiment results of an example participant.} (A) Control experiment, no diffusion version. The variance in the $x$-axis position of the disk at the moment of response is shown. Participants are instructed to press the correspondent key (right/left) as soon as the disk reaches either of the visible red lines that are 500 pixels away from the center. The $x$-axis represents the horizontal axis of the screen with 1920 pixels. The red symmetrical lines are the visual boundaries that the participants see throughout the whole trial. The number of observations of different positions of the disk at the response moment is shown in the $y$-axis . (B) Control experiment, $Y$-axis diffusion version. The variance in the $x$-axis position of the disk at the moment of response is shown. All participants show similar behavior. See Appendix~\ref{sec:addexp} for experimental details.
\label{figS1}}
\end{figure}

To further investigate the relation between variance and stimuli, we perform the set of control experiments explained in Sec.~\ref{sec:addexp} with decision-free response tasks.
In the no-diffusion (Fig.~\ref{figS1}A) and $Y$-axis diffusion (Fig.~\ref{figS1}B) versions of the control experiment, we observe substantially less variance in the $x$ positions of the responses.
Note that the stimuli in these two versions do not involve diffusion along the $x$ axis and therefore are fully deterministic along $x$.
The observed variance thus may come from two sources: (i) the motor implementation in pressing the key and/or (ii) the spatiotemporal discreteness of the stimuli.

\begin{figure}[!htbp]
\centering
\includegraphics[scale=0.4]{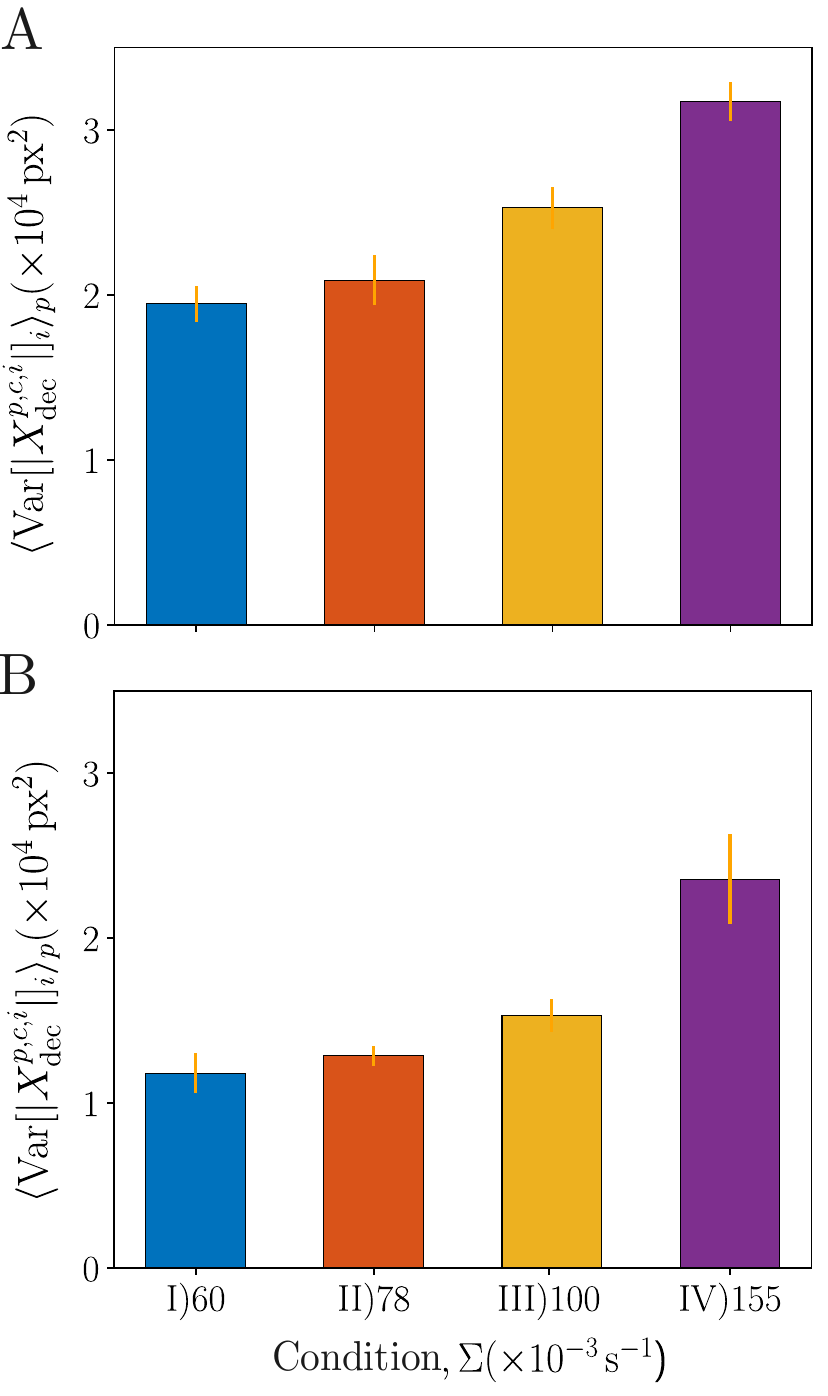}
\caption{\textbf{Variance in the $X_{\rm dec}$.} (A) The variance of $X_{\rm dec}$ averaged over participants across different experimental conditions in Experiment A. The variance in the decision threshold increases with entropy production rate. (B) Experimental results from the control experiment original parameters version. Participants are instructed to press the decision key as soon as the disk reaches visible red lines on the screen, representing Wald's SPRT-like fixed thresholds. Even though overall variance is decreased as opposed to Experiment A in panel A, there is still some variance, and it varies across conditions in a similar pattern as in panel A.}
\label{S6}
\end{figure}

We also conduct the original parameter version of the control experiment, where the same parameter values for the stimuli are used as Experiment A. 
Fig.~\ref{S6}A shows the variance in the decision thresholds of Experiment A. On the other hand, Fig.~\ref{S6}B depicts the variance of the responses in the original parameter version of the control experiment. 
The behavior of the variance as a function of the entropy production rate is similar in both experiments. 
In both cases, the increase in the variance may be explained by (i) the higher instantaneous jumps and (ii)  less predictability in the stimuli because of the higher entropy production rate.
Importantly, the variance in Experiment A is greater than in the control experiment. 
So far in the different versions of control experiments, we discuss the potential non-decisional sources of the variance in decision thresholds. The difference between Figs.~\ref{S6}A and ~\ref{S6}B suggests that part of the variance in thresholds in the main decision task stems from cognitive decision processes.
These cognitive decision processes that produce the variance in the experimental data may be explained by the EIM. As we have discussed in the Main Text, Fig.~\ref{fig5} shows the predictability power of the EIM.  

\section{Cross-validation of mean decision times}
\label{sec:cross_validation}

\begin{figure}[!htbp]
\centering
\includegraphics[scale=0.38]{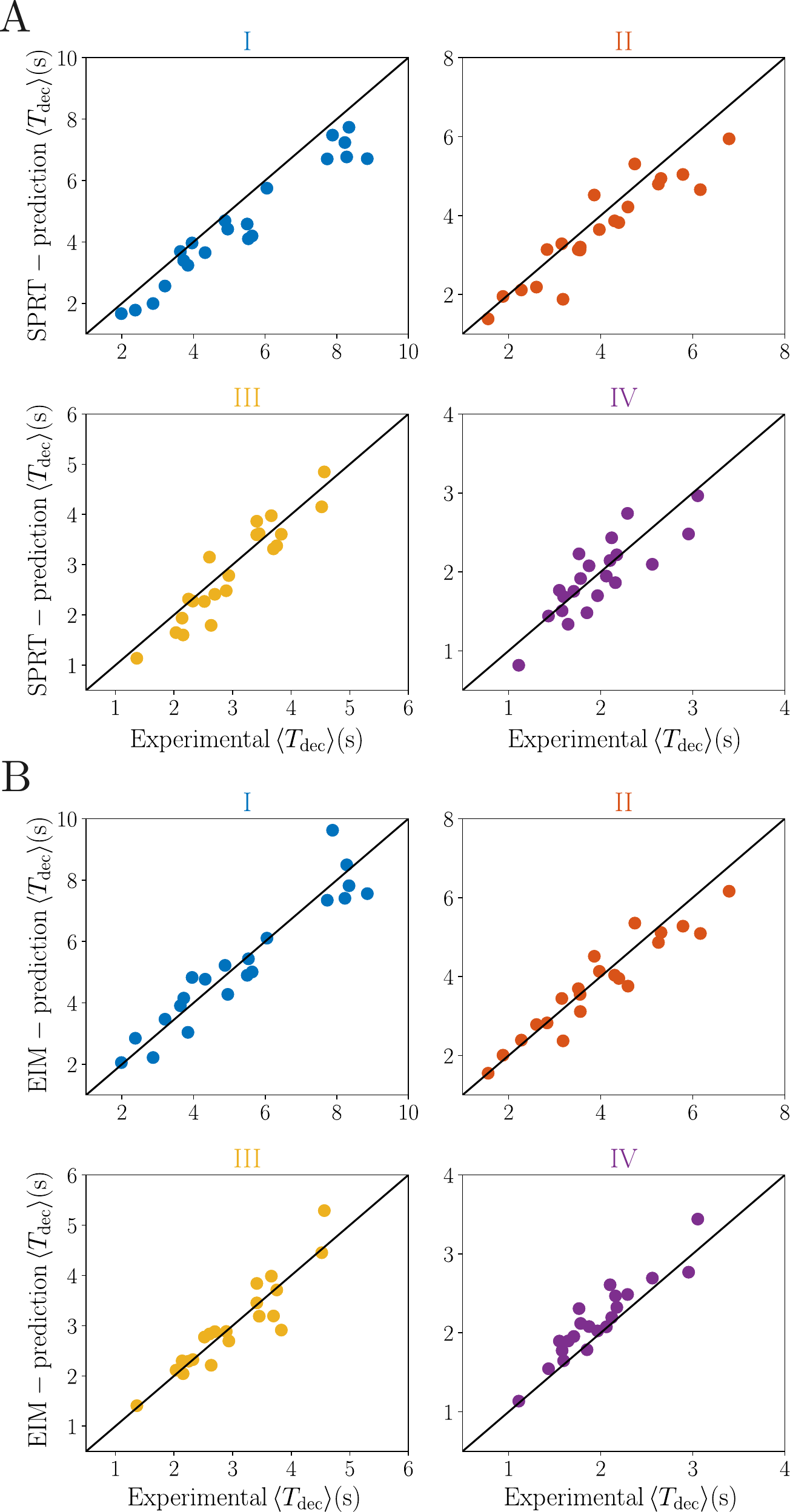}
\caption{\textbf{Cross validation of the model predictions of $T_{\rm dec}$ for (A) SPRT, and (B) EIM in Experiment A.} We find that the average experimental decision time of the training data accurately matches the predicted decision time of the test data for both (A) SPRT and (B) EIM. Each data point represents participants and the lines the best prediction.}\label{figS5}
\end{figure}

In order to test the robustness of the decision thresholds in predicting the decision times, we perform a cross-validation analysis in Experiment A for each participant in each condition. We divide the trial-by-trial data into two equal-sized segments at random (training and test) and use the mean decision thresholds of the training data to predict $T_{\rm dec}$ for the test data. Note that for the case of EIM to obtain the decision threshold, we use $\tau_m^\star$ in the training set trials. We find that the experimental mean decision time of the training data accurately matches the predicted mean decision time of the test data for both SPRT (see Fig.~\ref{figS5}A) and EIM (see Fig.~\ref{figS5}B). This suggests that knowledge of decision thresholds of a particular data set may be used to predict decision times of unseen data. 

\bibliographystyle{apsrev4-1}
\clearpage 

\end{document}